\documentclass{article}

\usepackage[tbtags]{amsmath}
\usepackage{amsfonts,amssymb}

\let\ln=\log
\newtheorem{proposition}{Proposition}
\newtheorem{theorem}{Theorem}

\newtheorem{lemma}{Lemma}

\newtheorem{corollary}{Corollary}

\newtheorem{gyph}{Conjecture}
\newtheorem{remark}{Remark}
\newtheorem{definition}{Definition}
\newtheorem{task}{Problem}

\newtheorem{example}{Example}

\newtheorem{step}{Step}
\newtheorem{demo}{Proof}

\newcommand\ptl{\partial}
\newcommand\tl{\tilde}
\newcommand\wtl{\widetilde}
\newcommand\vep{\varepsilon}
\newcommand\Bu{\mathbf{u}}
\newcommand\Bv{\mathbf{v}}
\newcommand\BZ{\mathbf{Z}}
\newcommand\cH{\mathcal{H}}
\newcommand\cL{\mathcal{L}}
\newcommand\rg{\mathrm{reg}}

\begin{document}

\centerline{{\LARGE \textbf {Integrable $(2{+}1)$-dimensional systems of hydrodynamic
type}}}
\bigskip
\bigskip
{\bf A.V. Odesskii ${}^{1}$,  V.V. Sokolov
${}^{2}$}
\\ [2ex] {\footnotesize
${}^{1}$  Brock University, St. Catharines, Ontario, Canada, e-mail:
aodesski@brocku.ca
\\
${}^{2}$  Landau Institute for Theoretical Physics, RAS, Moscow,
Russia, e-mail:
sokolov@itp.ac.ru
\\}

\bigskip

\begin{abstract}
We describe the results that have so far been obtained in the classification
problem for integrable $(2{+}1)$-dimensional systems of hydrodynamic type.
The Gibbons--Tsarev {\rm(}GT{\/\rm)} systems are most fundamental here. A
whole class of integrable $(2{+}1)$-dimensional models is related to each
such system. We present the known GT systems related to algebraic curves of
genus $g=0$ and $g=1$ and also a new GT system corresponding to algebraic
curves of genus $g=2$. We construct a wide class of integrable models
generated by the simplest GT system, which was not considered previously
because it is ``trivial."
\end{abstract}


\newpage
\tableofcontents

\section{Introduction}
\label{sec1}

Integrable systems play an important role in both mathematics and physics.
Unfortunately, a rigorous universal definition of integrability applicable
to differential equations of all types is lacking. Different views of the
integrability problem were described, for example, in~\cite{1}--\cite{4}.

Here, we consider systems of the form
\begin{equation}
\sum_{j=1}^n a_{ij}(\Bu)\frac{\ptl u_j}{\ptl t}+
\sum_{j=1}^n b_{ij}(\Bu)\frac{\ptl u_j}{\ptl y}+
\sum_{j=1}^n c_{ij}(\Bu)\frac{\ptl u_j}{\ptl x}=0,\qquad
\Bu=(u_1,\dots,u_n),
\label{1}
\end{equation}
where $i=1,\dots,n+k$ and $k\ge0$, and also equations that are reducible to
the above form by substitutions, for instance, quasilinear equations
\begin{equation}
A_1 Z_{tt}+A_2 Z_{yt}+A_3 Z_{xt}+A_4 Z_{yy}+A_5 Z_{xy}+A_6Z_{xx}=0,
\label{2}
\end{equation}
where $A_i=A_i(Z_x,Z_y,Z_t)$, and equations of the form
\begin{equation}
F(Z_{tt},Z_{yt},Z_{xt},Z_{yy},Z_{xy},Z_{xx})=0.
\label{3}
\end{equation}
Here and hereafter, we use the notation $Z_x=\ptl Z/\ptl x$, etc. To include
equations and systems of types~\eqref{2} and~\eqref{3} in our consideration,
we must allow the number of equations in system~\eqref{1} to exceed the
number of unknowns, i.e., we must be able to consider the case $k>0$.

The existence of a dispersionless zero-curvature
representation~\cite{5},~\cite{6} (also see~\cite{7}) or (which is the same)
a {\sl pseudo-potential representation} is a fundamental property of
integrable systems~\eqref{1}. This means that system~\eqref{1} is equivalent
to the compatibility conditions for the pair of the Hamilton--Jacobi
equations of the form
\begin{equation}
\Phi_y=A(\Phi_x,u_1,\dots,u_n),\qquad
\Phi_t=B(\Phi_x,u_1,\dots,u_n),
\label{4}
\end{equation}
where $\Phi(x,y,t)$ is a scalar function. As an example, we consider the
system
\begin{equation}
u_y=v_x,\qquad v_y=u_t-uu_x.
\label{5}
\end{equation}
We can easily verify that it admits the pseudopotential representation
\begin{equation}
\Phi_y=\frac{\Phi_x^2}2+u,\qquad\Phi_t=\frac{\Phi_x^3}3+u\Phi_x+v.
\label{6}
\end{equation}
Eliminating the unknown $v$, we can write system~\eqref{5} as the equation
$$
u_{yy}=(u_t-uu_x)_x,
$$
which is known as the dispersionless Kadomtsev--Petviashvili (KP) equation
or as the Khokhlov--Zabolots\-kaya equation.

For some integrable models, the functions $A$ and $B$ in the pseudopotential
representation depend on an additional spectral parameter $\lambda$. An
example is the dispersionless Hirota equation
\begin{equation}
a_1Z_xZ_{yt}+a_2Z_yZ_{xt}+a_3Z_tZ_{xy}=0,\qquad a_1+a_2+a_3=0,
\label{7}
\end{equation}
which has the pseudopotential representation
$$
\Phi_y=-\frac{a_2\lambda}{a_1\lambda+a_3}\frac{Z_y}{Z_x}\Phi_x,\qquad
\Phi_t=\lambda\frac{Z_t}{Z_x}\Phi_x.
$$
As in the case of integrable equations of the Korteveg--de~Vries (KdV)
equation type, the dependence on $\lambda$ can be polynomial, rational, and
so on. Assuming the existence of a pseudopotential representation with a
prescribed $\lambda$-dependence, we can construct examples of integrable
systems~\eqref{1}. For small $n$, we can also classify systems admitting
pseudopotential representations without the spectral parameter. But the very
existence of a pseudopotential representation cannot serve as a universal
integrability criterion for systems~\eqref{1} because various types of
pseudopotential representations exist.

The integrability of system~\eqref{1} is often related to its representation
as the commutation condition for a pair of vector
fields~\cite{8}--\cite{11}, which may or may not depend on the spectral
parameter $\lambda$. If a pseudopotential representation exists, then these
vector fields are Hamiltonian. Then, apparently, we cannot select a
constructively described class of vector fields covering all the known
examples and provide a rigorous universal definition of integrability based
on the commutation of vector fields.

In our view, we currently have only one property of integrable
systems~\eqref{1} that can be taken as a universal constructive
integrability criterion, which means that we not only can verify this
property for a given system but also can use it to construct new integrable
models and to provide a complete classification of these systems in lower
dimensions. This integrability criterion is that system~\eqref{1} has a
sufficient number of the so-called hydrodynamic reductions. The notion of
a hydrodynamic reduction arose in~\cite{12}, where the Benney chain
reductions were found. It turned out that $N$-component hydrodynamic
reductions can be described in terms of the compatible overdetermined
Gibbons--Tsarev (GT) system
\begin{equation}
\ptl_ip_j=\frac{\ptl_i u}{p_i-p_j},\qquad
\ptl_i\ptl_ju=\frac{2\ptl_iu\,\ptl_ju}{(p_i-p_j)^2},\qquad
i,j=1,\dots,N,\quad i\ne j.
\label{8}
\end{equation}
Here, the unknown functions $p_1,\dots,p_N$, and $u$ depend on some
variables $r^1,\dots,r^N$ and $\ptl_i=\ptl/\ptl r^i$. The same
system~\eqref{8} appears when describing hydrodynamic reductions of the
Khokhlov--Zabolotskaya equation.

It was mentioned in the pioneering papers~\cite{13}--\cite{15} that the
presence of hydrodynamic reductions can be effectively used to classify
integrable systems~\eqref{1}. Below, we briefly list the main classification
results. We describe the hydrodynamic reduction method itself in
Sec.~\ref{sec2}.

The problem of classifying integrable systems of the form
\begin{equation}
v_t+av_x+pv_y+qw_y=0,\qquad w_t+bw_x+rv_y+sw_y=0
\label{9}
\end{equation}
was considered in~\cite{14}. A system of 16 nonlinear partial differential
equations for the unknown coefficients $a(v,w)$, $b(v,w)$, $p(v,w)$,
$q(v,w)$, $r(v,w)$, and $s(v,w)$, which was equivalent to the existence of
hydrodynamic reductions, was written there. Because the obtained system for
the coefficients is in involution, we can use it to study general properties
of integrable systems~\eqref{9}. In particular, it was proved in~\cite{14}
that any integrable system~\eqref{9} admits a pseudopotential representation
and that the space of conservation laws of hydrodynamic type is
three-dimensional. The authors found many particular solutions of the system
of equations for the coefficients and thus constructed new examples of
integrable systems~\eqref{9}. But the authors did not construct the general
solution explicitly.

We can easily verify that for $n=l=2$, general system~\eqref{1} is reducible
to canonical form~\eqref{9} by a coordinate change of the form
\begin{equation}
v=\psi_1(u_1,u_2),\qquad w=\psi_2(u_1,u_2).
\label{10}
\end{equation}
But fixing canonical form~\eqref{9} destroys the group $GL(3)$ that acts on
the space of integrable systems~\eqref{1} by changing the independent
coordinates:
\begin{equation}
\bar t=k_{11}t+k_{12}x+k_{13}y,~
\bar y=k_{21}t+k_{22}x+k_{23}y,~\bar x=k_{31}t+k_{32}x+k_{33}y.
\label{11}
\end{equation}
Reducing the action of this group to its simplest form, Odesskii~\cite{16}
found another canonical form in which the system of equations for the
coefficients can be solved explicitly in terms of generalized hypergeometric
functions.

A special subclass of integrable equations~\eqref{2} with a Lagrangian of
the form $\cL(Z_t,Z_x,Z_y)$ was considered in~\cite{17},~\cite{18}.
Necessary and sufficient conditions for the Lagrangian integrability were
obtained in~\cite{17} in the form of an overdetermined system of partial
differential equations for the function $\cL$. General properties of
integrable Lagrangians and new examples were also constructed. A new
integrable Lagrangian $\cL=Z_tZ_xZ_y$ found there is the most interesting.
General integrable Lagrangians were described in~\cite{18} in terms of
modular forms.

The general classification problem for integrable equations~\eqref{2} was
considered in~\cite{19}, where an overdetermined system in involution for
the coefficients $A_i(Z_x,Z_y,Z_t)$ (which is equivalent to the existence of
hydrodynamic reductions) was written, general properties of integrable
equations~\eqref{2} were investigated, and new examples of these equations
were found. The general solution of this overdetermined system was found
in~\cite{20}. It depends on five arbitrary constants and can be expressed in
terms of generalized hypergeometric functions~\cite{21} of three variables
after an appropriate coordinate change. We note that in addition to
depending on the five essential parameters, the set of integrable
equations~\eqref{2} depends on 15 more parameters generated by the action of
the total linear group of transformations of the parameters $Z$, $x$, $t$,
and $y$, which is the invariance group for equation class~\eqref{2}.

Necessary and sufficient conditions for the integrability of equations of
form~\eqref{3} were obtained in~\cite{22}. It turned out that the function
$F$ does not contain essential parameters. Moreover, interesting classes of
particular solutions of the system for the function $F$ were found
in~\cite{22}. The general solution was constructed in~\cite{20} in terms of
the generalized hypergeometric function.

The classification problem for infinite integrable divergent chains of the form
\begin{equation}
u_{1t}=F_1(u_1,u_2)_x,~ u_{2t}=F_2(u_1,u_2,u_3)_x,\dots,
u_{it}=F_i(u_1,u_2,\ldots,u_{i+1})_x,\dots
\label{12}
\end{equation}
was considered in~\cite{23}--\cite{25}. The integrability was understood
in~\cite{23} as the existence of a commuting flow of the form
\begin{equation}
u_{1y}=G_1(u_1,u_2,u_3)_x,~ u_{2y}=G_2(u_1,u_2,u_3,u_4)_x,\dots,
u_{iy}=G_i(u_1,u_2,\dots,u_{i+2})_x,\dots
\label{13}
\end{equation}
A necessary condition for the existence of hydrodynamic reductions (the
vanishing of the Haantjes tensor) was used in~\cite{25} to produce the
system of equations for the functions $F_1$ and $F_2$ that must be satisfied
for the chain to be integrable.

In the above classification problems, in addition to having hydrodynamic
reductions, ``general position" integrable models have pseudopotential
representation~\eqref{4} without the spectral parameter. All the
pseudopotentials arising here are related to some rational curves. Krichever
constructed~\cite{26},~\cite{6} a class of pseudopotentials with arbitrarily
large $n$ related to algebraic curves of an arbitrary genus $n$. It can be
verified that the set of Krichever pseudopotentials corresponding to a
rational curve does not include all the examples found when classifying
system~\eqref{9} for $n=2$.

The definition of the pseudopotential integrability and the complete
description of integrable pseudopotentials for  $n=1$ were given in terms of
hydrodynamic reductions in~\cite{27}. The answer was formulated in terms of
hypergeometric functions.

Numerous examples related to rational curves lead to the problem of
constructing a class of pseudopotentials with arbitrary $n$ that contains
all these examples (possibly as reductions). Such a pseudopotential class
was constructed in~\cite{16},~\cite{20}. The result was expressed in terms
of a special class of generalized hypergeometric functions of $n$ variables
and contained $n{+}2$ essential parameters. When all these parameters are
integers, the result can be expressed in terms of elementary functions. The
problem of explicitly describing reductions of these potentials remains
open. The obtained results were generalized to the case of an elliptic curve
in~\cite{28}. Generalized elliptic hypergeometric functions appear in the
answer in~\cite{29}; the class of elliptic hypergeometric functions there is
apparently new.

Here, our presentation is based on the most fundamental notion of the
hydrodynamic reduction method: the GT systems. Although such systems
appeared previously~\cite{30}--\cite{34}, a rigorous definition of the class
of such systems, their equivalences, etc.~(see Sec.~\ref{sec3}), is
apparently still lacking in the literature. We present the known GT systems
related to algebraic curves of genus $g=0$ and $g=1$ and also a new GT
system related to algebraic curves of genus $g=2$. A whole class of
integrable $(2{+}1)$-dimensional models is related to each of the GT
systems. In Sec.~\ref{sec4}, we describe the results in~\cite{20},~\cite{28}
in terms of the GT system. In Sec.~\ref{sec5}, we construct a new broad
class of integrable models generated by the simplest GT system, which was
not considered previously because it is ``trivial." These systems have a
pseudopotential representation that depends on the spectral parameter
rationally.

In Sec.~\ref{sec6}, we share our experience acquired during several years of
work with systems~\eqref{1}. We devote Appendices~\ref{pr1}--\ref{pr4} to
conservation laws, pseudopotentials, multidimensional integrable
systems~\eqref{1}, and integrable chains. Appendix~\ref{pr1} is important
for understanding the relation between integrable systems~\eqref{1} and
integrable models of the type of Eqs.~\eqref{2} and~\eqref{3}. We describe
the relation between the GT systems and the pseudopotential representation
without the spectral parameter in Appendix~\ref{pr2}, which is important
both theoretically and technically. Examples of multidimensional integrable
systems of hydrodynamic type are in Appendix~\ref{pr3}, where we also
briefly discuss how to generalize the GT-system notion to the
multidimensional case. We devote Appendix~\ref{pr4} to the problem of
classifying arbitrary (not necessarily divergent) hydrodynamic-type chains
with the unit shift. Moreover, we present examples of integrable
multidimensional chains in Appendix~\ref{pr4}. We list unsolved problems in
Appendix~\ref{pr5}.

\section{Hydrodynamic reduction method}
\label{sec2}

It is well known that integrable equations of the KdV equation type
\begin{equation}
u_t=F\biggl(u,\frac{\ptl u}{\ptl x},\dots,\frac{\ptl^n u}{\ptl x^n}\biggr)
\label{14}
\end{equation}
have families of explicit solutions depending on arbitrary constants
$c_1,\dots,c_N$ for any $N$. All these finite-gap and solitonic solutions
can be constructed using the so-called ODE reductions. We call a pair of
compatible $N$-component systems of ODEs
\begin{equation}
r_x^i=f^i(r^1,\dots,r^N),\qquad r_t^i=g^i(r^1,\dots,r^N),\quad i=1,\dots,N,
\label{15}
\end{equation}
the $N$-component ODE reduction of Eq.~\eqref{14} if we have a function
$U(r^1,\dots,r^N)$ such that $u=U(r^1(x,t),$ $\dots,r^N(x,t))$ satisfies
Eq.~\eqref{14} for any solution $r^1(x,t),\dots,r^N(x,t)$ of
system~\eqref{15}. Clearly, a solution $u$ depends on $N$ arbitrary
parameters, which are initial values for system~\eqref{15} at the general
position point. We can take the existence of special classes of the ODE
reductions with an arbitrary $N$ as an integrability criterion for
Eq.~\eqref{14}. For instance, we can assume that Eq.~\eqref{14} admits a
sequence of differential constraints of the form
$$
\frac{\ptl^m u}{\ptl x^m}=G_m\biggl(u,\frac{\ptl u}{\ptl x},\dots,
\frac{\ptl^{m-1} u}{\ptl x^{m-1}}\biggr)
$$
with an arbitrary $m$. Clearly, we can rewrite Eq.~\eqref{14} and such a
differential constraint as a pair of compatible dynamical systems in the
variables $x$ and $t$. Another example of an ODE reduction is the Dubrovin
equations for zeros of the Baker--Akhiezer function in the KdV equation
theory. But much more effective and constructive integrability criteria
exist in the two-dimensional case. For example, we can take the existence of
higher local commuting flows (higher infinitesimal symmetries) or the
existence of higher conservation laws (see~\cite{35} and the references
therein) as such a criterion.

Higher local symmetries and/or local conservation laws are absent if the
number of independent variables is $d>2$ (see~\cite{36} for generalization
of the symmetry approach to the case of nonlocal symmetries). The existence
of $N$-component reductions can be considered one of the most prospective
approaches for seeking new integrable models in this situation. We note that
instead of the ODE reductions, we must there take some compatible systems of
partial differential equations of dimension not higher than $d-1$ as
reductions.

Such an approach was applied to systems of form~\eqref{1} in~\cite{13},
where pairs of compatible semi-Hamiltonian (see~\eqref{18} below) systems of
hydrodynamic type
\begin{equation}
r^i_t=\lambda^i(r^1,\dots,r^N)r^i_x,\qquad
r^i_y=\mu^i(r^1,\dots,r^N)r^i_x,\quad i=1,\dots,N,
\label{16}
\end{equation}
were taken as reductions. By the definition of the reduction, the
corresponding solutions of system~\eqref{1} are provided by some functions
$u^i(r^1,\dots, r^N)$, $i=1,\dots,n$, that transform any solution
of system~\eqref{16} into solution~\eqref{1}. Such solutions describe the
interaction of $N$ plane waves in hydrodynamics. They are sometimes called
the $N$-phase solutions.

Clearly, the generalized solution of a compatible overdetermined system of
form~\eqref{16} contains $N$ arbitrary functions of one variable. It turns
out that the functions $\lambda^i$ and $\mu^i$ in reduction~\eqref{16} may
contain additional functions of one variable as functional parameters; the
number of these functions does not exceed $N$. The existence of hydrodynamic
reductions~\eqref{16} locally parameterized by $N$ functions of one variable
with arbitrary $N$ was taken to be an integrability criterion for systems of
form~\eqref{1} in~\cite{13}. The corresponding $N$-phase solutions depend on
$2N$ arbitrary functions of one variable.

The approach based on hydrodynamic reductions is universal. This means that
all the currently known integrable systems~\eqref{1} admit hydrodynamic
reductions. All the notions in this approach can be rigorously defined (see
below). It was demonstrated in~\cite{13} that the existence of hydrodynamic
reductions can be verified algorithmically and can be effectively used to
classify integrable cases.

We recall the definitions and basic notions of the hydrodynamic reduction
method. Although they essentially coincide with those in~\cite{13}, we
introduce some technical improvements to make the definitions more rigorous.

The geometric theory of integrable $(1{+}1)$-dimensional hydrodynamic-type
systems of the form
\begin{equation}
r^i_t=\lambda^i(r^1,\dots,r^N)\,r^i_x,\quad i=1,\dots,N,
\label{17}
\end{equation}
was described in~\cite{37}. It was shown in~\cite{38} that these systems can
be integrated using the generalized hodograph method.

\begin{definition}
\label{def1}
System~\eqref{17} is said to be {\sl semi-Hamiltonian} if we have the
relations
\begin{equation}
\ptl_j\,\frac{\ptl_i\lambda^m}{\lambda^i-\lambda^m}=
\ptl_i\,\frac{\ptl_j\lambda^m}{\lambda^j-\lambda^m},\quad i\ne j\ne m,\qquad
\ptl_i=\frac{\ptl}{\ptl r^i}.
\label{18}
\end{equation}
Semi-Hamiltonian systems have an infinite number of commuting flows and
conservation laws of hydrodynamic type~\cite{38},~\cite{39}. The main
geometric object related to semi-Hamiltonian system~\eqref{17} is the
diagonal metric $g_{kk}$, $k=1,\dots,N$, where
\begin{equation}
\frac12\ptl_i\ln g_{kk}=\frac{\ptl_i\lambda^k}{\lambda^i-\lambda^k},\quad
i\ne k,
\label{19}
\end{equation}
which is called the {\sl metric associated with}~\eqref{17}.
\end{definition}

\begin{definition}
\label{def2}
A hydrodynamic reduction of system~\eqref{1} is governed by a pair of
compatible semi-Hamiltonian systems of hydrodynamic type~\eqref{16} and by
the functions $u_1(r^1,\dots,r^N),\dots,u_n(r^1,\dots,r^N)$ such that the
functions
\begin{equation}
u_1=u_1(r^1,\dots,r^N),\qquad\dots,\qquad u_n=u_n(r^1,\dots,r^N)
\label{20}
\end{equation}
satisfy system~\eqref{1} for any solution of system~\eqref{16}.
Following~\cite{13}--\cite{15}, we say that system~\eqref{1} is {\sl
integrable} if it admits the maximum possible number of hydrodynamic
reductions. Namely, substituting~\eqref{20} in~\eqref{1}, eliminating
derivatives with respect to $t$ and $y$ using~\eqref{16}, and equating the
coefficients of $r^l_x$ to zero, we obtain
\begin{align}
&\sum_{j=1}^na_{ij}(\Bu)\lambda^l\ptl_lu_j+
\sum_{j=1}^nb_{ij}(\Bu)\mu^l\ptl_l u_j+\sum_{j=1}^nc_{ij}(\Bu)\ptl_l u_j=0,
\label{21}
\\[2mm]
&i=1,\dots,n+k,\quad l=1,\dots,N.
\nonumber
\end{align}
\end{definition}

For each $l$, we have a linear overdetermined system for $n$ unknowns
$\ptl_lu_1,\dots,\ptl_lu_n$ whose coefficients are independent of $l$.
Because this system must have a nontrivial solution, all its $n{\times}n$
minors must vanish, which results in the system of algebraic equations for
$\lambda^l$ and $\mu^l$, which is the same for all $l$. We assume that this
system is equivalent to a single algebraic equation
\begin{equation}
P(\lambda^l,\mu^l)=0
\label{22}
\end{equation}
(otherwise, $\lambda^l$ and $\mu^l$ are fixed, and we then have just a
finite number of hydrodynamic reductions). Equation~\eqref{22} determines
the so-called algebraic {\sl dispersion curve}. Let $p$ be a coordinate on
this curve. Then Eq.~\eqref{22} is equivalent to the equations
$$
\lambda^l=F(p_l,u_1,\dots,u_n),\qquad\mu^l=G(p_l,u_1,\dots,u_n)
$$
for some functions $F$ and $G$. It was assumed in~\cite{13} that linear
system~\eqref{21} has a unique solution up to proportionality at a general
value of $p_l$. Solving this system, we obtain
\begin{equation}
\ptl_iu_m=g_m(p_i,u_1,\dots,u_n)\ptl_iu_1,\quad
m=2,\dots,n,\quad i=1,\dots,N,
\label{23}
\end{equation}
for some functions $g_m$. We rewrite~\eqref{16} in the form
\begin{equation}
r^i_t=F(p_i,u_1,\dots,u_n)r^i_x,\qquad r^i_y=G(p_i,u_1,\dots,u_n)r^i_x,\quad
i=1,\dots,N.
\label{24}
\end{equation}
It is easy to see that the compatibility conditions for~\eqref{24} are
\begin{equation}
\frac{\ptl_iF(p_j)}{F(p_i)-F(p_j)}=\frac{\ptl_iG(p_j)}{G(p_i)-G(p_j)}
\label{25}
\end{equation}
(where we omit the arguments $u_1,\dots,u_n$ of $F$ and $G$).
We express $\ptl_ip_j$ from~\eqref{25}:
\begin{equation}
\ptl_i p_j=f(p_i,p_j,u_1,\dots,u_n)\ptl_i u_1,\quad
i\ne j,\quad i,j=1,\dots,N.
\label{26}
\end{equation}
Finally, the compatibility conditions $\ptl_i\ptl_ju_m=\ptl_j\ptl_iu_m$
result in the equation
\begin{equation}
\ptl_i\ptl_ju_1=h(p_i,p_j,u_1,\dots,u_n)\ptl_iu_1\ptl_ju_1,\quad
i\ne j,\quad i,j=1,\dots,N.
\label{27}
\end{equation}
Collecting Eqs.~\eqref{23},~\eqref{26}, and~\eqref{27} together, we obtain
the system of equations
\begin{equation}
\begin{alignedat}2
&\ptl_ip_j=f(p_i,p_j,u_1,\dots,u_n)\ptl_iu_1,&\quad
&i\ne j,\quad i,j=1,\dots,N,
\\[2mm]
&\ptl_iu_m=g_m(p_i,u_1,\dots,u_n)\ptl_iu_1,&\quad
&m=2,\dots,n,\quad i=1,\dots,N,
\\[2mm]
&\ptl_i\ptl_ju_1=h(p_i,p_j,u_1,\dots,u_n)\ptl_i u_1\ptl_j u_1,&\quad
&i\ne j,\quad i,j=1,\dots,N.
\end{alignedat}
\label{28}
\end{equation}
Here, $p_1,\dots,p_N$ and $u_1,\dots,u_n$ are functions of $r^1,\dots,r^N$,
$N\ge3$, and $\ptl_i=\ptl/\ptl r^i$. Because system~\eqref{1} must have the
maximum number of reductions by assumption, system~\eqref{28} must be in
involution (i.e., it must be completely compatible). This means that the
compatibility conditions $\ptl_k\ptl_jp_j=\ptl_j\ptl_kp_j$ and
$\ptl_k\ptl_i\ptl_ju_1=\ptl_i\ptl_k\ptl_ju_1$ are satisfied by virtue of
system~\eqref{28}. Solutions of system~\eqref{28} therefore depend on $2N$
functions of one variable. We can take these functions to be boundary values
of the Goursat problem $u_1(0,\dots,0,r^i,0,\dots,0)$ and
$p_i(0,\dots,0,r^i,0,\dots,0)$, where $i=1,\dots,N.$

\section{Gibbons--Tsarev systems}
\label{sec3}

In the preceding section, we showed that a system of form~\eqref{28} is
related to each system~\eqref{1} that has the maximum set of hydrodynamic
reductions.

\begin{definition}
\label{def3}
A compatible system of form~\eqref{28} is called an $n$-{\sl field GT
system.}
\end{definition}

\begin{definition}
\label{def4}
Two GT systems are said to be {\sl equivalent} if they are related by a
transformation of the form
\begin{alignat}2
&p_i\to\lambda(p_i,u_1,\dots,u_n),&\quad&i=1,\dots,N,
\label{29}
\\[2mm]
&u_m\to\mu_m(u_1,\dots,u_n),&\quad&m=1,\dots,n.
\label{30}
\end{alignat}
\end{definition}

\begin{definition}
\label{def5}
Transformation~\eqref{29},~\eqref{30} is called a {\sl GT-system
automorphism} if it transforms this system into itself.
\end{definition}

It is easy to see that system~\eqref{28} is compatible if and only if the
functions $f$, $g_k$, and $h$ satisfy a system of functional equations
independent of $N$. For example, for the one-field GT system
\begin{equation}
\ptl_i p_j=f(p_i,p_j,u)\ptl_iu,\qquad
\ptl_i\ptl_ju=h(p_i,p_j,u)\ptl_iu\ptl_ju,\quad i\ne j,\quad i,j=1,\dots,N,
\label{31}
\end{equation}
these equations are
\begin{equation}
\begin{aligned}
f_u(p_2,p_3,u)-f_u(p_1,p_3,u)+f_{p_2}(p_2,p_3,u)f(p_1,p_2,u)-
f_{p_1}(p_1,p_3,u)f(p_2,p_1,u)+{}&
\\[2mm]
{}+f_{p_3}(p_2,p_3,u) f(p_1,p_3,u)-f_{p_3}(p_1,p_3,u)f(p_2,p_3,u)+{}&
\\[2mm]
{}+f(p_2,p_3,u)h(p_1,p_2,u)-f(p_1,p_3,u)h(p_1,p_2,u)&=0,
\\[2mm]
h_u(p_2,p_3,u)-h_u(p_1,p_3,u)+h_{p_2}(p_2,p_3,u)f(p_1,p_2,u)-
h_{p_1}(p_1,p_3,u)f(p_2,p_1,u)+{}&
\\[2mm]
{}+h_{p_3}(p_2,p_3,u)f(p_1,p_3,u)-h_{p_3}(p_1,p_3,u)f(p_2,p_3,u)+{}&
\\[2mm]
{}+h(p_2,p_3,u)h(p_1,p_2,u)-h(p_1,p_3,u)h(p_1,p_2,u)&=0.
\end{aligned}
\label{32}
\end{equation}

\medskip

{\bf3.1.~Examples of GT systems}

\begin{example}
\label{ex1}
Let $a_0$, $a_1$, and $a_2$ be arbitrary constants. Then the equations
\begin{equation*}
\begin{aligned}
&\ptl_ip_j=\frac{a_2p_j^2+a_1p_j+a_0}{p_i-p_j}\ptl_iu\qquad
\\[2mm]
&\ptl_i\ptl_ju=\frac{2a_2p_ip_j+a_1(p_i+p_j)+2a_0}{(p_i-p_j)^2}
\ptl_iu\ptl_ju,
\end{aligned}\quad i,j=1,\dots,N,\quad i\ne j,
\end{equation*}
determine a one-field GT system. The original GT system (see
Sec.~\ref{sec1}) corresponds to the case where $a_2=a_1=0$ and $a_0=1$. In
the general position case, we can use a linear transformation of the
variables $p_i$ to reduce the system to the form
\begin{equation}
\begin{aligned}
&\ptl_ip_j=\frac{p_j(p_j-1)}{p_i-p_j}\ptl_iu,
\\[2mm]
&\ptl_i\ptl_ju=\frac{2p_ip_j-p_i-p_j}{(p_i-p_j)^2}\ptl_iu\ptl_ju,
\end{aligned}\quad i,j=1,\dots,N,\quad i\ne j.
\label{33}
\end{equation}
\end{example}

\begin{example}
\label{ex2}
Let $P(x)=a_3x^3+a_2x^2+a_1x+ a_0$. Then the system of equations
\begin{equation}
\begin{aligned}
&\ptl_ip_j=\frac{P(p_j)(p_i-u)}{P(u)(p_i-p_j)}\ptl_i u,
\\[2mm]
&\ptl_i\ptl_ju=\frac{K_2(p_i,p_j)u^2+K_1(p_i,p_j)u+
K_0(p_i,p_j)}{P(u)(p_i-p_j)^2}\ptl_iu\ptl_ju,
\end{aligned}\quad i,j=1,\dots,N,\quad i\ne j,
\label{34}
\end{equation}
where
\begin{align*}
&K_2(p_i,p_j)=2a_3(p_i-p_j)^2,
\\[2mm]
&K_1(p_i,p_j)=-a_3(p_i^2p_j+p_ip_j^2)+
a_2(p_i^2+p_j^2-4p_ip_j)-a_1(p_i+p_j)-2a_0,
\\[2mm]
&K_0(p_i,p_j)=2a_3p_i^2p_j^2+a_2(p_i^2p_j+p_ip_j^2)+
a_1(p_i^2+p_j^2)+a_0(p_i+p_j),
\end{align*}
is a one-field GT system. Using transformations of the form
$$
u\to\frac{au+b}{cu+d},\qquad p_i\to\frac{ap_i+b}{cp_i+d},
$$
we can reduce the polynomial $P$ to one of the three canonical forms
$P(x)=x(x-1)$, $P(x)=x$, or $P(x)=1$. We note that in the general position
case where $P(x)=x(x-1)$, the system has an automorphism group $S_4$
interchanging the points 0, 1, $\infty$, and $u$ and generated by the
elements
$$
\sigma_1{:}\;u\to1-u,\quad p_i\to1-p_i,\qquad
\sigma_2{:}\;u\to\frac u{u-1},\quad p_i\to\frac{p_i}{p_i-1},
$$
and
$$
\sigma_3{:}\;u\to1-u,\quad p_i\to\frac{(1-u) p_i}{p_i-u}.
$$
\end{example}

\begin{example}
\label{ex3}
We set
$$
\theta(z,\tau)=\sum_{\alpha\in\mathbb Z}(-1)^{\alpha}
e^{2\pi i(\alpha z+\alpha(\alpha-1)\tau/2)},\qquad
\rho(z,\tau)=\frac{\theta_z(z,\tau)}{\theta(z,\tau)}.
$$
Then the system
\begin{equation}
\ptl_{\alpha}p_{\beta}=\frac1{2\pi i}(\rho(p_{\alpha}-p_{\beta})-
\rho(p_{\alpha}))\ptl_{\alpha}\tau,\qquad
\ptl_{\alpha}\ptl_{\beta}\tau=-\frac1{\pi i}\rho'(p_{\alpha}-p_{\beta})
\ptl_{\alpha}\tau\ptl_{\beta}\tau,
\label{35}
\end{equation}
where $\alpha,\beta=1,\dots,N$, $\alpha\ne\beta$, is a one-field GT system.
The field $u$ here is the modular parameter $\tau$. We note that the same
system completed with some equations with $\alpha=\beta$ appeared
in~\cite{32} in relation to a very different problem.
\end{example}

\begin{example}
\label{ex4}
Let $u$, $v$, and $w$ be the coordinates on the moduli space of curves
\begin{equation}
y^2=x(x-1)(x-u)(x-v)(x-w)
\label{36}
\end{equation}
of genus $g=2$. Then the formulas
\begin{align*}
&\ptl_iv=\frac{v(v-1)(p_i-u)}{u(u-1)(p_i-v)}\ptl_iu,
\\[2mm]
&\ptl_iw=\frac{w(w-1)(p_i-u)}{u(u-1)(p_i-w)}\ptl_iu,
\\[2mm]
&\ptl_ip_j=\frac{p_j(p_j-1)(p_i-u)(p_i-v)(p_i-w)+y(p_i)y(p_j)}
{u(u-1)(p_i-v)(p_i-w)(p_i-p_j)}\ptl_iu,
\\[2mm]
&\ptl_i\ptl_ju=\biggl(\frac{u(p_i+p_j)+(u-1)(p_i^2+p_j^2)+(p_i+p_j-4u)p_ip_j}
{u(u-1)(p_i-p_j)^2}+{}
\\[2mm]
&\phantom{\ptl_i\ptl_ju={}}+\frac{(2vw-(p_i+p_j)(v+w)+2p_ip_j)y(p_i)y(p_j)}
{u(u-1)(p_i-v)(p_i-w)(p_j-v)(p_j-w)(p_i-p_j)^2}\biggr)\ptl_iu\ptl_ju
\end{align*}
determine a three-field GT system.
\end{example}

The systems in Examples~\ref{ex3} and~\ref{ex4} are related to the moduli
spaces of curves of the respective genera $g=1$ and $g=2$. If we
parameterize elliptic curves using the formula $y^2=x(x-1)(x-u)$ instead of
the modular parameter, then we can rewrite the system in Example~\ref{ex3}
in a form similar to the system in Example~\ref{ex4}.

The functions $f$ and $h$ have poles at $p_i=p_j$ in
Examples~\ref{ex1}--\ref{ex4}. But there are GT systems holomorphic at
$p_i=p_j$.

\begin{example}
\label{ex5}
The formulas
\begin{equation}
\ptl_ip_j=0,\qquad\ptl_iu_m=g_m(p_i)\ptl_iu_1,\qquad\ptl_i\ptl_ju_1=0
\label{37}
\end{equation}
define an $n$-field GT system for any $n$ and for arbitrary functions
$g_m(x)$.
\end{example}

\medskip

{\bf3.2.~GT-system extensions}

\begin{definition}
\label{def6}
An {\sl extension of GT system}~\eqref{28} using the field functions
$u_{n+1},\dots,u_{n+m}$ is an auxiliary system of the form
\begin{equation}
\ptl_iu_k=g_k(p_i,u_1,\dots,u_{n+m})\ptl_iu_1,\quad
k=n+1,\dots,n+m,\quad i=1,\dots,N,
\label{38}
\end{equation}
such that Eqs.~\eqref{28} and~\eqref{38} are compatible.
\end{definition}

{\bf Example 1a} (continuation of Example~\ref{ex1}){\bf.} System~\eqref{33}
admits the one-field extensions
\begin{align}
&\ptl_iv=\frac{v(v-1)}{p_i-v}\ptl_iu ,
\label{39}
\\[2mm]
&\ptl_iv=\biggl(\frac1{p_i^2}+\frac v{p_i}\biggr)\ptl_iu,
\label{40}
\\[2mm]
&\ptl_iv=\biggl(\frac1{(p_i-1)^2}-\frac v{p_i-1}\biggr)\ptl_iu,
\label{41}
\\[2mm]
&\ptl_iv=\biggl(\frac{c_1}{p_i-1}+\frac{c_2}{p_i}+
(c_3e^u-c_4e^{2u})p_i+c_4e^{2u}p_i^2\biggr)\ptl_iu,
\label{42}
\end{align}
where $c_i$ are arbitrary constants.

\medskip

Describing all extensions for a given GT system is a complicated problem.
But extensions of type~\eqref{39} always exist and can be constructed in a
uniform way. Namely, it turns out that if we complete system~\eqref{28} with
the equation
$$
\ptl_iu_{n+1}=f(p_i,u_{n+1},u_1,\dots,u_n)\ptl_iu_1
$$
(the function $f$ is the same as in~\eqref{28}), then the obtained
($n+1$)-field system remains compatible. We call this procedure the {\sl
regular extension.} For example, the regular extension in the case of
Example~\ref{ex1} results in the additional equation
$$
\ptl_iv=\frac{a_2v^2+a_1v+a_0}{p_i-v}\ptl_iu.
$$
We note that if we take $v$ as a new field variable, then the first formula
in Example~\ref{ex1} becomes
$$
\ptl_ip_j=\frac{(a_2p_j^2+a_1p_j+a_0)(p_i-v)}
{(a_2v^2+a_1v+a_0)(p_i-p_j)}\ptl_iv,\quad i,j=1,\dots,N,\quad i\ne j.
$$
We see that we can use this trick to obtain the canonical forms of the GT
system in Example~\ref{ex2} from the corresponding GT systems in
Example~\ref{ex1}.

Clearly, we can apply the regular extension procedure repeatedly. For
example, the $n$-fold regular extension of GT system~\eqref{8} is
$$
\ptl_ip_j=\frac1{p_i-p_j}\ptl_iw,\qquad
\ptl_i\ptl_jw=\frac2{(p_i-p_j)^2}\ptl_iw\ptl_jw,\qquad
\ptl_iu_k=\frac{\ptl_iw}{p_i-u_k},\quad k=1,\dots,n.
$$
For this system, we can construct degenerations corresponding to
merging some fields. In particular, merging all the fields $u_i\to w$,
$i=1,\dots,n$, results in the GT system
\begin{align*}
&\ptl_ip_j=\frac1{p_i-p_j}\ptl_iw_1,\qquad
\ptl_i\ptl_jw_1=\frac2{(p_i-p_j)^2}\ptl_iw_1\ptl_jw_1,
\\[2mm]
&\ptl_iw_j=\biggl(p_i^{j-1}-\sum_{k=1}^{j-2}ku_kp_i^{j-k-2}\biggr)\ptl_iw_1.
\end{align*}
This extension at $n=1$ exactly coincides with the GT system for
dispersionless KP equation~\eqref{5} (see Example~\ref{ex6} below).

We note that we cannot obtain the GT system in Example~\ref{ex4} from a
system with fewer fields using a regular extension procedure.

\section{From GT systems to integrable models}
\label{sec4}

Constructing compatible families of systems of form~\eqref{24} is
our next step on the path from GT systems to integrable models~\eqref{1}.

\medskip

{\bf4.1. Families of $(1{+}1)$-dimensional hydrodynamic-type systems
associated with a GT system}

\begin{definition}
A family of semi-Hamiltonian systems of hydrodynamic type
\begin{equation}
r^i_t=F(p_i,u_1,\dots,u_n)r^i_x
\label{43}
\end{equation}
parameterized by solutions $p_i(r^1,\dots,r^N)$, $i=1,\dots,N$,
$u_j(r^1,\dots,r^N)$, $j=1,\dots,n$, of GT system~\eqref{28} is called a
{\sl family of $(1{+}1)$-dimensional systems associated with
system~\eqref{28}}.
\end{definition}

Two $(1{+}1)$-dimensional systems~\eqref{24} are compatible if and only if
the functions $F$ and $G$ satisfy functional equation~\eqref{25}. This
equation has trivial solutions $F$ and $G=c_1F+c_2$, where $F$ is an
arbitrary function and $c_i$ are arbitrary constants.

Not every GT system admits nontrivial solutions $F$ and $G$ of
Eq.~\eqref{25} depending essentially on the field functions $u_i$. For
example, such solutions exist only under a special choice of the functions
$g_m$ in Example~\ref{ex5}.

We consider one-field GT systems~\eqref{31}. In accordance with~\eqref{25},
we must then have functions of two variables $F(x,u)$ and $G(x,u)$ such that
\begin{equation}
\frac{f(x,y,u)\frac{\ptl F}{\ptl y}(y,u)+
\frac{\ptl F}{\ptl u}(y,u)}{F(x,u)-F(y,u)}=
\frac{f(x,y,u)\frac{\ptl G}{\ptl y}(y,u)+
\frac{\ptl G}{\ptl u}(y,u)}{G(x,u)-G(y,u)}.
\label{44}
\end{equation}

\begin{proposition}
\label{prop1}
Every one-field GT system admitting a nontrivial solution $F(x,u)$, $G(x,u)$
of Eq.~\eqref{44} is equivalent to a system in either Example~\ref{ex1} or
Example~\ref{ex2}.
\end{proposition}

The following results on constructing solutions of functional
equation~\eqref{44} for $n$-field systems with arbitrary $n$ were obtained
in~\cite{20}, \cite{28}.

\medskip

{\bf Example 1b} (continuation of Example~\ref{ex1}){\bf.}
An $n$-fold application of the regular extension procedure to GT
system~\eqref{33} with a quadratic polynomial of general position results in
the GT system
\begin{align}
&\ptl_ip_j=\frac{p_j(p_j-1)}{p_i-p_j}\ptl_iw,\qquad
\ptl_i\ptl_jw=\frac{2p_ip_j-p_i-p_j}{(p_i-p_j)^2}\ptl_iw\,\ptl_jw,
\label{45}
\\[2mm]
&i,j=1,\dots,N,\quad i\ne j,
\nonumber
\\[2mm]
&\ptl_iu_j=\frac{u_j(u_j-1)\ptl_iw}{p_i-u_j},\quad
i=1,\dots,N,\quad j=1,\dots,n.
\label{46}
\end{align}
We consider the compatible overdetermined system of partial differential
equations
\begin{align}
\frac{\ptl^2 h}{\ptl u_j\ptl u_k}={}&\frac{s_j}{u_j-u_k}
\frac{\ptl h}{\ptl u_k}+\frac{s_k}{u_k-u_j}\frac{\ptl h}{\ptl u_j},\quad
i,j=1,\dots,n,\quad j\ne k,
\label{47}
\\[2mm]
\frac{\ptl^2h}{\ptl u_j\ptl u_j}={}&-\biggl(1+\sum_{k=1}^{n+2}s_k\biggr)
\frac{s_j}{u_j(u_j-1)}h+\frac{s_j}{u_j (u_j-1)}
\sum_{\substack{k=1,\\k\ne j\hfill}}^n
\frac{u_k(u_k-1)}{u_k-u_j}\frac{\ptl h}{\ptl u_k}+{}
\nonumber
\\[2mm]
&{}+\Biggl(\,\sum_{\substack{k=1,\\k\ne j\hfill}}^n\frac{s_k}{u_j-u_k}+
\frac{s_j+s_{n+1}}{u_j}+
\frac{s_j+s_{n+2}}{u_j-1}\Biggr)\frac{\ptl h}{\ptl u_j},
\label{48}
\end{align}
where $s_1,\dots,s_{n+2}\in\mathbb C$ are arbitrary parameters. Solutions of
this system belong to the class of generalized hypergeometric
functions~\cite{21}. We can easily see that the vector space $\cH$ of all
the solutions of this system has the dimension $n+1$. For any $h\in\cH$, we
set
\begin{equation}
S(h,p)=\sum_{i=1}^nu_i(u_i-1)(p-u_1)\cdots\hat{i}\cdots(p-u_n)h_{u_i}+
\biggl(1+\sum_{i=1}^{n+2}s_i\biggr)(p-u_1)\cdots(p-u_n)h,
\label{49}
\end{equation}
where $\hat{i}$ means that we omit the corresponding multiplier in the
product. The function $S$ is obviously a polynomial of degree $n$ in $p$.

Let $h_1$, $h_2$, and $h_3$ be linearly independent functions in $\cH$.
Then the functions
\begin{equation}
F=\frac{S(h_1,p)}{S(h_3,p)},\qquad G=\frac{S(h_2,p)}{S(h_3,p)}
\label{50}
\end{equation}
satisfy functional equation~\eqref{25}.

Moreover, we have solutions of Eq.~\eqref{25} of form~\eqref{50} for which
$S$ has a degree less than $n$. In this case, the polynomial $S$ is to be
defined in a more complicate way. Namely, we fix $k$ linearly independent
functions $h_1,\dots,h_k\in \cH$. Then for each $h\in\cH$, the polynomial
$S(h,p)$ is given by the formula
\begin{equation}
S(h,p)=\frac1{\Delta}\sum^{n-k+1}_{i=1}
u_i(u_i-1)(p-u_1)\cdots\hat{i}\cdots(p-u_{n-k+1})\Delta_i(h),
\label{51}
\end{equation}
where $h\in \cH$ and
\begin{align*}
&\Delta=\det\begin{pmatrix}
h_1&h_2&\cdots&h_k\\[1mm]
h_{1,u_{n-k+2}}&h_{2,u_{n-k+2}}&\cdots&h_{k,u_{n-k+2}}\\[1mm]
\vdots&\vdots&\ddots&\vdots\\[1mm]
h_{1,u_n}&h_{2,u_n}&\dots&h_{k,u_n}\end{pmatrix},
\\[2mm]
&\Delta_i(h)=\det\begin{pmatrix}
h&h_1&h_2&\cdots&h_k\\[1mm]
h_{u_i}&h_{1,u_i}&h_{2,u_i}&\cdots&h_{k,u_i}\\[1mm]
h_{u_{n-k+2}}&h_{1,u_{n-k+2}}&h_{2,u_{n-k+2}}&\cdots&h_{k,u_{n-k+2}}\\[1mm]
\vdots&\vdots&\vdots&\ddots&\vdots\\[1mm]
h_{u_n}&h_{1,u_n}&h_{2,u_n}&\dots&h_{k,u_n}\end{pmatrix}.
\end{align*}
Clearly, $S(h,p)$ is a polynomial of degree $n-k$ in $p$. We note that
linear transformations $h_i\to c_{i1}h_1+\dots+c_{ik}h_k$ and $h\to
h+d_1h_1+\dots+d_kh_k$ with constant coefficients $c_{ij}$ and $d_i$ do not
change $S(h,p)$.

There are solutions of functional equation~\eqref{25} for GT
system~\eqref{45},~\eqref{46} different from those constructed above. For
example, for $n=2$, we have the solution
$$
F=\frac{(p-u_1)(p-u_2)}{p (u_1-u_2)},\qquad
G=\frac{(p-u_1)(p-u_2)}{(p-1) (u_1-u_2)}.
$$
Apparently, these solutions are related to singular rational algebraic
curves.

\medskip

{\bf Example 3a} (continuation of Example~\ref{ex3}){\bf.}
A regular $n$-field extension of GT system~\eqref{35} is given by the
formulas
\begin{equation}
\ptl_{\alpha}u_{\beta}=\frac1{2\pi i}(\rho(p_{\alpha}-u_{\beta})-
\rho(p_{\alpha}))\ptl_{\alpha}\tau,\quad\beta=1,\dots,n.
\label{52}
\end{equation}
A nontrivial solution of functional equation~\eqref{25} is given by
formulas~\eqref{50} with
$$
S(h, p)=\sum_{\alpha=1}^n\frac{\theta(u_{\alpha})\theta(p-u_{\alpha}-\eta)}
{\theta(u_{\alpha}+\eta)\theta(p-u_{\alpha})}h_{u_{\alpha}}-
(s_1+\dots+s_n)\frac{\theta'(0)\theta(p-\eta)}{\theta(\eta)\theta(p)}h.
$$
Here, $\eta=s_1u_1+\dots+s_nu_n+r\tau+\eta_0$, where $s_1,\dots,s_n,r,\eta_0$
are arbitrary constants and $h(u_1,\dots,u_n,\tau)$ is a solution of the
elliptic hypergeometric system
\begin{align*}
&h_{u_{\alpha}u_{\beta}}=s_{\beta}(\rho(u_{\beta}-u_{\alpha})+
\rho(u_{\alpha}+\eta)-\rho(u_{\beta})-\rho(\eta))h_{u_{\alpha}}+{}
\\[2mm]
&\phantom{h_{u_{\alpha}u_{\beta}}={}}+s_{\alpha}(\rho(u_{\alpha}-u_{\beta})+
\rho(u_{\beta}+\eta)-\rho(u_{\alpha})-\rho(\eta))h_{u_{\beta}},
\\[2mm]
&h_{u_{\alpha}u_{\alpha}}=s_{\alpha}\sum_{\beta\ne\alpha}(\rho(u_{\alpha})+
\rho(\eta)-\rho(u_{\alpha}-u_{\beta})-\rho(u_{\beta}+\eta))h_{u_{\beta}}+{}
\\[2mm]
&\phantom{h_{u_{\alpha}u_{\alpha}}={}}+\biggl(\,\sum_{\beta\ne\alpha}
s_{\beta}\rho(u_{\alpha}-u_{\beta})+(s_{\alpha}+1)\rho(u_{\alpha}+\eta)+
s_{\alpha}\rho(-\eta)+{}
\\[2mm]
&\phantom{h_{u_{\alpha}u_{\alpha}}={}}+(s_0-s_{\alpha}-1)\rho(u_{\alpha})+
2\pi ir\biggr)h_{u_{\alpha}}-s_0s_{\alpha}(\rho'(u_{\alpha})-\rho'(\eta))h,
\\[2mm]
&h_{\tau}=\frac1{2\pi i}\sum_{\beta}(\rho(u_{\beta}+\eta)-
\rho(\eta))h_{u_{\beta}}-\frac{s_0}{2\pi i}\rho'(\eta)h.
\end{align*}
We can show that the solution space of this system has the dimension $n+1$.

We construct elliptic analogues of solutions~\eqref{50},~\eqref{51} as
follows. We fix $k$ linearly independent solutions $h_1,\dots,h_k$ of the
elliptic hypergeometric system. For any solution $h$, we then have
\begin{equation}
S(h,p)=\frac1{\Delta}\biggl(\,\sum_{\alpha=1}^{n-k}
\frac{\theta(u_{\alpha})\theta(p-u_{\alpha}-\eta)}
{\theta(u_{\alpha}+\eta)\theta(p-u_{\alpha})}\Delta_{\alpha}(h)-
(s_1+\dots+s_n)\,\frac{\theta'(0)\theta(p-\eta)}
{\theta(\eta)\theta(p)}\,\Delta_0(h)\biggr),
\label{53}
\end{equation}
where
\begin{align*}
&\Delta=\det\begin{pmatrix}
h_{1,u_{n-k+1}}&\cdots&h_{k,u_{n-k+1}}\\[1mm]
\vdots&\ddots&\vdots\\[1mm]
h_{1,u_n}&\cdots&h_{k,u_n}\end{pmatrix},
\\[2mm]
&\Delta_{\alpha}(h)=\det\begin{pmatrix}
h_{u_{\alpha}}&h_{1,u_{\alpha}}&\cdots&h_{k,u_{\alpha}}\\[1mm]
h_{u_{n-k+1}}&h_{1,u_{n-k+1}}&\cdots&h_{k,u_{n-k+1}}\\[1mm]
\vdots&\vdots&\ddots&\vdots\\[1mm]
h_{u_n}&h_{1,u_n}&\dots&h_{k,u_n}\end{pmatrix},
\\[2mm]
&\Delta_0(h)=\det\begin{pmatrix}
h&h_1&\cdots&h_k\\[1mm]
h_{u_{n-k+1}}&h_{1,u_{n-k+1}}&\cdots&h_{k,u_{n-k+1}}\\[1mm]
\vdots&\vdots&\ddots&\vdots\\[1mm]
h_{u_n}&h_{1,u_n}&\cdots&h_{k,u_n}\end{pmatrix}.
\end{align*}

\medskip

{\bf4.2. Integrable three-dimensional systems generated by a pair of
compatible families.}
Given GT system~\eqref{28} and a pair of compatible families~\eqref{24}
(i.e., a solution $F$, $G$ of functional equation~\eqref{25}), we can
construct the corresponding system~\eqref{1} by a simple calculation.

\begin{lemma}
\label{lem1}
We consider the linear space $V$ of functions of the variable $p$ generated
by $3n$ functions
$$
\{F(p,u_1,\dots,u_n)g_j(p,u_1,\dots,u_n),\;
G(p,u_1,\dots,u_n)g_j(p,u_1,\dots,u_n),\;
g_j(p,u_1,\dots,u_n);\;j=1,\dots,n\bigr\}.
$$
Here, $g_1=1$ by definition. The system of form~\eqref{1} with hydrodynamic
reductions~\eqref{24} then consists of $n{+}k$ equations if and only if the
dimension of $V$ is $2n-k$. The coefficients of system~\eqref{1} are
determined by the relations
\begin{align}
&\sum_{j=1}^n \bigl(a_{ij}(\Bu)F(p,u_1,\dots,u_n)+
b_{ij}(\Bu)G(p,u_1,\dots,u_n)+c_{ij}(\Bu)\bigr)g_j(p,u_1,\dots,u_n)=0,
\label{54}
\\[2mm]
&i=1,\dots,n+k.
\nonumber
\end{align}
\end{lemma}

\begin{corollary}
If the functions $F$ and $G$ are determined by formulas of type~\eqref{50},
where the expression $S(h,p)$ is linear in $h$ and $h_1$, $h_2$, and $h_3$
are arbitrary elements of a vector space, then the linear transformation
$\bar h_i=\sum_{j=1}^3 k_{ij}h_j$ corresponds to transformation~\eqref{11}
in the corresponding system~\eqref{1}.
\end{corollary}

\begin{example}
\label{ex6}
We consider the extension $\ptl_iv=p_i\ptl_iu$ of GT system~\eqref{8}. The
corresponding functional equation~\eqref{44} has the solution $F=p^2+u$,
$G=p$. Moreover, we have $g_1=1$ and $g_2=p$. The space generated by the set
of functions $\{p^2+u,\;p^3+up,\;p,\;p^2,\;1,\;p\}$ has the dimension
$\alpha=4$. The matrices $A$, $B$, and $C$ composed of the unknown
coefficients of system~\eqref{1} are determined up to simultaneous
multiplication by an arbitrary matrix from the left. We can easily verify
that using such a transformation, we can reduce any solution of the system
of eight equations, which is equivalent to relations~\eqref{54}, to
$a_{11}=1$, $a_{12}=a_{21}=a_{22}=0$, $b_{11}=b_{22}=0$, $b_{12}=-1$,
$b_{21}=1$, $c_{11}=-u$, $c_{12}=c_{21}=0$, and $c_{22}=-1$. This solution
corresponds to system~\eqref{5}.
\end{example}

An analogous calculation for~\eqref{45},~\eqref{46},~\eqref{49},
and~\eqref{50} results in the system for $u_1,\dots,u_n$:
\begin{align}
\sum_{i\ne j}(h_{q,u_j}h_{r,u_i}-h_{r,u_j}h_{q,u_i})
\frac{u_j(u_j-1)u_{i,t_s}-u_i(u_i-1)u_{j,t_s}}{u_j-u_i}+{}\qquad&
\nonumber
\\[2mm]
{}+\sigma(h_qh_{r,u_j}-h_rh_{q,u_j})u_{j,t_s}+{}\qquad&
\nonumber
\\[2mm]
{}+\sum_{i\ne j}(h_{r,u_j}h_{s,u_i}-h_{s,u_j}h_{r,u_i})
\frac{u_j(u_j-1)u_{i,t_q}-u_i(u_i-1)u_{j,t_q}}{u_j-u_i}+{}&
\nonumber
\\[2mm]
{}+\sigma(h_rh_{s,u_j}-h_sh_{r,u_j})u_{j,t_q}+{}&
\nonumber
\\[2mm]
{}+\sum_{i\ne j}(h_{s,u_j}h_{q,u_i}-h_{q,u_j}h_{s,u_i})
\frac{u_j(u_j-1)u_{i,t_r}-u_i(u_i-1)u_{j,t_r}}{u_j-u_i}+{}&
\nonumber
\\[2mm]
{}+\sigma(h_sh_{q,u_j}-h_qh_{s,u_j})u_{j,t_r}&=0,
\label{55}
\end{align}
where $j=1,\dots,n$ and $\sigma=1+s_1+\dotsb+s_{n+2}$. Here, the subscript
after the comma indicates the partial derivative with respect to the
corresponding argument, $q=0$, $r=1$, $s=2$ and $t_0=x$, $t_1=t$, $t_2=y$ in
expressions like $h_{q,u_j}$ and $u_{i,t_q}$.

\begin{remark}
Let $h_0,h_1,h_2,\dots,h_n$ be a basis in the space of solutions of
hypergeometric system~\eqref{47},~\eqref{48}. We associate the proper time
$t_i$ with each $h_i$. For each triple of pairwise distinct
$0\le q,r,s\le n$, formula~\eqref{55} defines the corresponding
three-dimensional system with the times $t_q$, $t_r$, and $t_s$. According
to~\cite{20}, all these systems are mutually compatible.
\end{remark}

The explicit form of systems related to the case $k>0$ can be found
in~\cite{20}. Systems generated by the elliptic GT system in
Example~\ref{ex3} were presented in~\cite{28}.

\section{Integrable weakly nonlinear $(2{+}1)$-dimensional systems}
\label{sec5}

The functions $f$ and $h$ have a pole on the diagonal $p_i=p_j$ in all
GT-system examples~\eqref{28} above. But we also have GT systems holomorphic
at $p_i=p_j$. In particular, such are the systems in Example~\ref{ex5}.

\begin{definition}
\label{def8}
Integrable system~\eqref{1} is said to be {\sl weakly nonlinear} if the
corresponding GT system is holomorphic on the diagonal $p_i=p_j$.
\end{definition}

The GT-system holomorphicity conditions are expressed as follows in terms of
the solution $F$, $G$ of functional equation~\eqref{25}. Writing~\eqref{25}
in the form
\begin{align*}
&\biggl(\ptl_ip_jF'(p_j)+\sum_{\alpha}F(p_j)_{u_{\alpha}}
\ptl_iu_{\alpha }\biggr)(G(p_i)-G(p_j))=\biggl(\ptl_ip_jG'(p_j)+
\sum_{\alpha}G(p_j)_{u_{\alpha}}\ptl_iu_{\alpha}\biggr)(F(p_i)-F(p_j)),
\end{align*}
where the prime denotes the derivative with respect to $p_j$, and expressing
$\ptl_ip_j$ from it, we find that this expression is holomorphic on the
diagonal if and only if
\begin{equation}
\sum_{\alpha}\bigl(F(p_i)_{u_{\alpha}}G'(p_i)-
G(p_i)_{u_{\alpha}}F'(p_i)\bigr)\ptl_iu_{\alpha}=0.
\label{56}
\end{equation}

\begin{proposition}
\label{prop2}
Let system~\eqref{1} be weakly nonlinear, and let $k=0$ {\rm(}i.e., the
number of unknowns is equal to the number of equations in the system{\rm).}
Then the matrix
$$
\tl A=(a_{11}A+a_{12}B+a_{13}C)^{-1}(a_{21}A+a_{22}B+a_{23}C)
$$
is linearly degenerate for all constant $a_{ij}$.
\end{proposition}

We recall that a matrix $P(u_1,\dots,u_n)$ is said to be linearly degenerate
if the derivative of any eigenvalue along the corresponding eigenvector
vanishes.

\begin{demo}
Using substitution~\eqref{29}, we reduce $G(p)$ to the form $G=p$.
Then~\eqref{56} becomes
\begin{equation}
\sum_\alpha F(p_i)_{u_{\alpha}}\ptl_iu_{\alpha}=0.
\label{57}
\end{equation}
An arbitrary change of variables~\eqref{11} brings system~\eqref{1} to the
form $\Bu_t=\tl A \Bu_x+\wtl B \Bu_y$, where
\begin{align*}
\tl A&=(a_{11}A+a_{12}B+a_{13}C)^{-1}(a_{21}A+a_{22}B+a_{23}C),
\\[2mm]
\wtl B&=(a_{11}A+a_{12}B+a_{13}C)^{-1}(a_{31}A+a_{32}B+a_{33}C).
\end{align*}
For $\mu^l=p_l$ and $\lambda^l=F(p_l)$, formula~\eqref{21} for this system
yields
$$
(\tl A+p_i\wtl B-F(p_i)\operatorname{Id})
(\ptl_i u_1,\dots,\ptl_i u_n)^{\mathrm T}=0.
$$
For this system, formula~\eqref{57} implies that the derivative of the
eigenvalue $F(p_i)$ along the eigenvector $(\ptl_i u_1,\dots,\ptl_i u_n)$
vanishes. The matrix $\tl A+p_i\wtl B$ is therefore linearly degenerate at
any value of the parameter $p_i$.
\end{demo}

The obtained result states that if the number of equations in
system~\eqref{1} is equal to the number of unknown functions, then every
two-dimensional system describing solutions $\Bu=\Bu(c_1x+c_2y+c_3t,
c_4x+c_5y+c_6t)$ of the traveling-wave type for weakly nonlinear
system~\eqref{1} is weakly nonlinear in the sense in~\cite{40}.

\begin{example}
\label{ex7}
We consider the two-component system (see~\cite{13})
\begin{equation}
\begin{pmatrix}v_t\\[1mm]w_t\end{pmatrix}+
\begin{pmatrix}a&0\\[1mm]0&b\end{pmatrix}
\begin{pmatrix}v_x\\[1mm]w_x\end{pmatrix}+
\begin{pmatrix}p&q\\[1mm]r&s\end{pmatrix}
\begin{pmatrix}v_y\\[1mm]w_y\end{pmatrix}=0,
\label{58}
\end{equation}
where
\begin{align*}
&a=w,\qquad b=v,\qquad r=\frac{P(w)}{w-v},\qquad q=\frac{P(v)}{v-w},
\\[2mm]
&s=\frac{P(v)}{w-v}+\frac13P'(v),\qquad p=\frac{P(w)}{v-w}+\frac13P'(w).
\end{align*}
Here, $P$ is an arbitrary third-degree polynomial. The corresponding GT
system
\begin{align*}
&\ptl_ip_j=\frac{P(w)}{(w-v)P(v)}p_j^2p_i+
\biggl(\frac1{w-v}+\frac{P'(v)}{P(v)}\biggr)p_jp_i-{}
\\[2mm]
&\phantom{\ptl_ip_j={}}-\biggl(\frac1{v-w}+\frac{P'(w)}{P(w)}\biggr)p_j-
\frac{P(v)}{(v-w)P(w)},
\\[2mm]
&\ptl_iv=p_i\ptl_i w,
\\[2mm]
&\ptl_i\ptl_jw=\biggl(\frac{P(w)}{(v-w)P(v)}p_ip_j+
\frac1{v-w}+\frac{P'(w)}{P(w)}\biggr)\ptl_iw\ptl_jw
\end{align*}
is polynomial in $p_i$ and $p_j$, and the initial $(2{+}1)$-dimensional
system is therefore weakly nonlinear. We can verify that this GT system is
equivalent to the GT system
$$
\ptl_ip_j=0,\qquad\ptl_iu_2=p_i\ptl_iu_1,\qquad\ptl_i\ptl_ju_1=0,
$$
which belongs to the class in Example~\ref{ex5}.
\end{example}

As shown in~\cite{13}, system~\eqref{9} has a linear pseudopotential
representation with the spectral parameter. This turns out to be a general
property of systems~\eqref{37} in Example~\ref{ex5}.

\begin{proposition}
\label{prop3}
Let $F(p,u_1,\dots,u_n)$ and $G(p,u_1,\dots,u_n)$ be a solution of
functional equation~\eqref{25} for a GT system of form~\eqref{37}. Then the
corresponding $(2{+}1)$-dimensional system admits the pseudopotential
representation
\begin{equation}
\psi_t=F(\lambda,u_1,\dots,u_n)\,\psi_x,\qquad
\psi_y=G(\lambda,u_1,\dots,u_n)\,\psi_x,
\label{59}
\end{equation}
where $\lambda$ is the spectral parameter.
\end{proposition}

\begin{demo}
From~\eqref{25} with the vanishing of $\ptl_ip_j$ taken into account, we
obtain
\begin{equation}
\label{60}
\frac{\sum_{\alpha}F(p_j)_{u_{\alpha}}\ptl_iu_{\alpha}}{F(p_i)-F(p_j)}=
\frac{\sum_{\alpha}G(p_j)_{u_{\alpha}}\ptl_iu_{\alpha}}{G(p_i)-G(p_j)}.
\end{equation}
On the other hand, the compatibility condition for system~\eqref{59} is
$F_y+FG_x=G_t+GF_x$, whence we obtain
$$
\sum_{\alpha}\bigl(F(\lambda)_{u_{\alpha}}u_{\alpha y}+
F(\lambda) G(\lambda)_{u_{\alpha}} u_{\alpha x}\bigr)=
\sum_{\alpha}\bigl(G(\lambda)_{u_{\alpha}}u_{\alpha t}+
G(\lambda) F(\lambda)_{u_{\alpha}} u_{\alpha x}\bigr).
$$
Substituting the expressions
$$
u_{\alpha y}=\sum_i\ptl_iu_{\alpha}r^i_y=\sum_i\ptl_iu_{\alpha}G(p_i)r^i_x,
\qquad u_{\alpha t}=\sum_i\ptl_iu_{\alpha}F(p_i)r^i_x
$$
in this equality and replacing $\lambda$ with $p_j$, we obtain~\eqref{60}.
\end{demo}

\subsection{The general position case}
\label{s51}
Using our observation on the equivalence between GT system~\eqref{58} and a
system of form~\eqref{37} with rational functions $g_m$, we now generalize
Example~\ref{ex7} to the case of arbitrary $n$ and $k$.

We consider an $(n{+}1)$-field GT system
\begin{equation}
\ptl_ip_j=0,\qquad\ptl_iu_m=
\frac{\lambda_m-\lambda_0}{p_i-\lambda_m}\ptl_iw,\qquad\ptl_i\ptl_jw=0,
\label{61}
\end{equation}
where $u_1,\dots,u_n$ and $w$ are the field functions and $\lambda_0,
\lambda_1,\dots,\lambda_n$ are the parameters. We let $\cH_n$ denote the
linear space of functions in the variables $u_1,\dots,u_n$ generated by the
elements $1,e^{u_1},\dots,e^{u_n}$. For any function $g=a_0+a_1e^{u_1}+
\dots+a_ne^{u_n}\in\cH_n$, we set
$$
S_n(g,p)=\frac{a_0}{p-\lambda_0}+\sum_{i=1}^n\frac{a_ie^{u_i}}{p-\lambda_i}.
$$
For any $k\in\mathbb N$ such that $0<k<n-1$, we fix the functions $h_1,
\dots,h_k\in\cH_n$, where $h_i=b_{i,0}+b_{i,1}e^{u_1}+\dots+_{i,n}e^{u_n}$,
and set
\begin{equation}
S_{n,k}(g,p)=\det\begin{pmatrix}
S_n(g,p)&S_n(h_1,p)&S_n(h_2,p)&\cdots&S_n(h_k,p)\\[1mm]
g&h_1&h_2&\cdots&h_k\\[1mm]
a_{n-k+2}&b_{1,n-k+2}&b_{2,n-k+2}&\cdots&b_{k,n-k+2}\\[1mm]
\vdots&\vdots&\vdots&\ddots&\vdots\\[1mm]
a_n&b_{1,n}&b_{2,n}&\cdots&b_{k,n}\end{pmatrix}.
\label{62}
\end{equation}
By definition, $S_{n,0}(g,p)=S_n(g,p)$.

\begin{theorem}
\label{t1}
Let $g_1$, $g_2$, and $g_3$ be linearly independent functions in $\cH_n$.
Then for any $0\le k<n-1$, the functions
\begin{equation}
F=\frac{S_{n,k}(g_1,p_i)}{S_{n,k}(g_3,p_i)},\qquad
G=\frac{S_{n,k}(g_2,p_i)}{S_{n,k}(g_3,p_i)}
\label{63}
\end{equation}
satisfy functional equation~\eqref{25}.
\end{theorem}

We now write the corresponding $(2{+}1)$-dimensional systems explicitly.
According to the lemma in Sec.~4.2, the equation
$$
\sum_{i=1}^n(A_iu_{i,t_1}+B_iu_{i,t_2}+C_iu_{i,x})=0
$$
belongs to the system if and only if the expression
$$
\sum_{i=1}^n\frac{\lambda_i-\lambda_0}{p-\lambda_i}
\bigl(A_iS_{n,k}(g_1,p)+B_iS_{n,k}(g_2,p)+C_iS_{n,k}(g_3,p)\bigr)
$$
vanishes identically. We first consider the case $k=0$. Let $g_i=a_{i,0}+
a_{i,1}e^{u_1}+\dots+a_{i,n}e^{u_n}$, $i=1,2,3$. Then the corresponding
$(2{+}1)$-dimensional system is
\begin{align}
\sum^n_{\substack{j=1,\\j\ne i\hfill}}(a_{2,i}a_{3,j}-a_{2,j}a_{3,i})e^{u_j}
\frac{u_{i,t_1}-u_{j,t_1}}{\lambda_i-\lambda_j}+(a_{2,i}a_{3,0}-a_{3,i}
a_{2,0})\frac{u_{i,t_1}}{\lambda_i-\lambda_0}+{}\qquad&
\nonumber
\\[2mm]
{}+\sum^n_{\substack{j=1,\\j\ne i\hfill}}(a_{3,i}a_{1,j}-a_{3,j}a_{1,i})
e^{u_j}\frac{u_{i,t_2}-u_{j,t_2}}{\lambda_i-\lambda_j}+(a_{3,i}a_{1,0}-
a_{1,i}a_{3,0})\frac{u_{i,t_2}}{\lambda_i-\lambda_0}+{}&
\nonumber
\\[2mm]
{}+\sum^n_{\substack{j=1,\\j\ne i\hfill}}(a_{1,i}a_{2,j}-a_{1,j}a_{2,i})
e^{u_j}\frac{u_{i,x}-u_{j,x}}{\lambda_i-\lambda_j}+(a_{1,i}a_{2,0}-
a_{2,i}a_{1,0})\frac{u_{i,x}}{\lambda_i-\lambda_0}&=0,
\label{64}
\end{align}
where $i=1,\dots,n$. In the simplest case, $n=2$, and we can use linear
transformations of the functions $g_i$ (see the corollary in Sec.~4.2) to
reduce the matrix $a_{ij}$ to the diagonal form $\operatorname{diag}\bigl(
(\lambda_1-\lambda_0)^{-1},(\lambda_0-\lambda_2)^{-1},(\lambda_2-
\lambda_1)^{-1}\bigr)$. System~\eqref{64} then takes the simple form
$$
u_y=e^v(v_x-u_x),\qquad v_t=e^u(u_x-v_x).
$$
The system in Example~\ref{ex7} reduces to this system upon some changes of
form~\eqref{11},~\eqref{10}. Introducing the potential
$$
Z_t=e^u,\qquad Z_y=-e^v,
$$
we can write this system as the single equation
$$
\frac{Z_{ty}}{Z_t Z_y}=\frac{Z_{xy}}{Z_y}+\frac{Z_{xt}}{Z_t}.
$$

If $k>0$, then the corresponding $(2{+}1)$-dimensional system contains two
groups of equations:
\begin{align*}
\sum^{n-k+1}_{\substack{j=1,\\j\ne i\hfill}}\bigl(\Delta_i(g_2)
\Delta_j(g_3)-\Delta_j(g_2)\Delta_i(g_3)\bigr)e^{u_j}
\frac{u_{i,t_1}-u_{j,t_1}}{\lambda_i-\lambda_j}+{}\qquad\qquad&
\\[2mm]
{}+\bigl(\Delta_i(g_2)\Delta_0(g_3)-\Delta_0(g_2)\Delta_i(g_3)\bigr)
\frac{u_{i,t_1}}{\lambda_i-\lambda_0}+{}&
\\[2mm]
{}+\sum^{n-k+1}_{\substack{j=1,\\j\ne i\hfill}}\bigl(\Delta_i(g_3)
\Delta_j(g_1)-\Delta_j(g_3)\Delta_i(g_1)\bigr)e^{u_j}
\frac{u_{i,t_2}-u_{j,t_2}}{\lambda_i-\lambda_j}+{}&
\\[2mm]
{}+\bigl(\Delta_i(g_3)\Delta_0(g_1)-\Delta_0(g_3)\Delta_i(g_1)\bigr)
\frac{u_{i,t_2}}{\lambda_i-\lambda_0}+{}&
\\[2mm]
{}+\sum^{n-k+1}_{\substack{j=1,\\j\ne i\hfill}}\bigl(\Delta_i(g_1)
\Delta_j(g_2)-\Delta_j(g_1)\Delta_i(g_2)\bigr)e^{u_j}
\frac{u_{i,x}-u_{j,x}}{\lambda_i-\lambda_j}+{}&
\\[2mm]
{}+\biggl(\Delta_i(g_1)\Delta_0(g_2)-\Delta_0(g_1)\Delta_i(g_2)\biggr)
\frac{u_{i,x}}{\lambda_i-\lambda_0}&=0,
\end{align*}
where $i=1,\dots,n-k$, and
\begin{align*}
&\sum_{j=1}^{n-k+1}e^{u_j}\Delta_j(g_r)u_{j,t_s}=
\sum_{j=1}^{n-k+1}e^{u_j}\Delta_j(g_s)u_{j,t_r},
\\[2mm]
&\sum_{j=1}^{n-k+1}\Delta_j(g_r)e^{u_j}
\frac{u_{i,t_s}-u_{j,t_s}}{\lambda_i-\lambda_j}+
\Delta_0(g_r)\frac{u_{i,t_s}}{\lambda_i-\lambda_0}=
\sum_{j=1}^{n-k+1}\Delta_j(g_s)e^{u_j}\frac{u_{i,t_r}-u_{j,t_r}}
{\lambda_i-\lambda_j}+\Delta_0(g_s)\frac{u_{i,t_r}}{\lambda_i-\lambda_0},
\end{align*}
where $i=n-k+2,\dots,n$. Here, $r,s=1,2,3$, $t_3=x$, and
\begin{align*}
&\Delta_j(g)=\det\begin{pmatrix}
g&h_1&h_2&\cdots&h_k\\[1mm]
a_j&b_{1,j}&b_{2,j}&\cdots&b_{k,j}\\[1mm]
a_{n-k+2}&b_{1,n-k+2}&b_{2,n-k+2}&\cdots&b_{k,n-k+2}\\[1mm]
\vdots&\vdots&\vdots&\ddots&\vdots\\[1mm]
a_n&b_{1,n}&b_{2,n}&\cdots&b_{k,n}\end{pmatrix},\quad j=1,\dots,n,
\\[2mm]
&g=a_0+a_1e^{u_1}+\dots+a_ne^{u_n},
\\[2mm]
&h_1=b_{1,0}+b_{1,1}e^{u_1}+\dots+b_{1,n}e^{u_n},\qquad\dots,\qquad
h_k=b_{k,0}+b_{k,1}e^{u_1}+\dots+b_{k,n}e^{u_n}.
\end{align*}
The first group contains $n{-}k$ equations, and the second group contains
$3k$ equations. There are exactly $k$ linear combinations of equations in
the second group that contain no derivatives of the $u_i$, $i=n-k+1,\dots,n$.
These linear combinations can be linearly expressed in terms of the equations
in the first group. The $(2{+}1)$-dimensional system therefore contains
$(n-k)+3k-k=n+k$ linearly independent equations.

The simplest nontrivial case is where $n=3$ and $k=1$. In
Appendix~\ref{pr1}, we show that the corresponding $(2{+}1)$-dimensional
system is equivalent to Eq.~\eqref{7}.

\subsection{Degeneracies}
\label{s52}
Gibbons--Tsarev system~\eqref{61} and formulas~\eqref{63} correspond to the
case of pairwise distinct parameters $\lambda_0,\dots,\lambda_n$. Here, we
consider some reductions of the construction in Sec.~\ref{s51} that
correspond to merging these parameters.

We define the polynomials $P_i(u_1,u_2,\dots)$ as the coefficients of the
Taylor expansion
$$
e^{\vep u_1+\vep^2u_2+\cdots}=1+P_1\vep+P_2\vep^2+\ldots.
$$
In particular,
$$
P_1(u_1,u_2,\dots)=u_1,\qquad P_2(u_1,u_2,\dots)=u_2+\frac12u_1^2,\qquad
P_3(u_1,u_2,\dots)=u_3+u_1u_2+\frac16u_1^3.
$$
We let $Q_k(\vep,u_1,u_2,\dots)$ denote the partial sums $1+\sum_{i=1}^kP_i
\vep^i$. We set $P_0=Q_0 =1$ by definition.

Let $\lambda_0,\dots,\lambda_l$ be pairwise distinct roots with the
respective multiplicities $n_0+1,n_1,\dots,n_l$. Clearly, $n_0+\dots+n_l=n$.
The corresponding $(n{+}1)$-field GT system with the field functions
$u_{0,1},\dots,u_{0,n_0},u_{1,1},\dots,$ $u_{1,n_1},\dots,u_{l,1},\dots,
u_{l,n_l},w$ is
\begin{equation}
\begin{aligned}
&\ptl_ip_j=0,\qquad\ptl_iu_{0,m}=\frac1{(p_i-\lambda_0)^m}\ptl_iw,
\\[2mm]
&\ptl_iu_{s,m}=\biggl(\frac{\lambda_s-\lambda_0}{(p_i-\lambda_s)^m}+
\frac1{(p_i-\lambda_s)^{m-1}}\biggr)\ptl_iw,\qquad\ptl_i\ptl_jw=0.
\end{aligned}
\label{65}
\end{equation}

In the case of GT system~\eqref{65}, the solutions $F,G$ of functional
equation~\eqref{25} are given by the formulas in the theorem above where the
functions $S_{n,k}$ are defined as follows. We let $\cH$ denote the vector
space generated by the functions $P_i(u_{0,1},u_{0,2},\dots)$, $i=0,\dots,
n_0$, and $e^{u_{s,1}}P_{i-1}(u_{s,1},u_{s,2},\dots)$, $s=1,\dots,l$,
$i=1,\dots,n_s$. For any element
$$
g=\sum_{i=0}^{n_0}a_{0,i}P_i(u_{0,1},u_{0,2},\dots)+\sum_{s=1}^l
\sum_{i=1}^{n_s}a_{s,i}e^{u_{s,1}}P_{i-1}(u_{s,1},u_{s,2},\dots)
$$
in this space, we set
$$
S_n(g,p)=\sum_{i=0}^{n_0}\frac{a_{0,i}Q_i(p-\lambda_0,u_{0,1},u_{0,2},\dots)}
{(p-\lambda_0)^{i+1}}+\sum_{s=1}^l\sum_{i=1}^{n_s}
\frac{a_{s,i}e^{u_{s,1}}Q_{i-1}(p-\lambda_s,u_{s,1},u_{s,2},\dots)}
{(p-\lambda_s)^i}.
$$
By definition, $S_{n,0}(g,p)=S_n(g,p)$.

In the case $k>0$, we fix the elements
$$
h_j=\sum_{i=0}^{n_0}b_{0,i}P_i(u_{0,1},u_{0,2},\dots)+\sum_{s=1}^l
\sum_{i=1}^{n_s}b_{s,i}e^{u_{s,1}}P_{i-1}(u_{s,1},u_{s,2},\dots),\quad
j=1,\dots,k,
$$
in $\cH$. Then
\begin{equation}
S_{n,k}(g,p)=\det\begin{pmatrix}
S_n(g,p)&S_n(h_1,p)&S_n(h_2,p)&\cdots&S_n(h_k,p)\\[1mm]
g&h_1&h_2&\cdots&h_k\\[1mm]
g_{\Bv_{n-k+2}}&h_{1,\Bv_{n-k+2}}&h_{2,\Bv_{n-k+2}}&\cdots
&h_{k,\Bv_{n-k+2}}\\[1mm]
\vdots&\vdots&\vdots&\ddots&\vdots\\[1mm]
g_{\Bv_n}&h_{1,\Bv_n}&h_{2,\Bv_n}&\cdots&h_{k,\Bv_n}\end{pmatrix},
\label{66}
\end{equation}
where $(\Bv_1,\dots,\Bv_n)=(u_{0,1},\dots,u_{0,n_0},u_{1,1},\dots,u_{1,n_1},
\dots,u_{l,1},\dots,u_{l,n_l})$. We can then easily reconstruct the explicit
form of the corresponding $(2{+}1)$-dimensional systems using the general
construction in the lemma (see Sec.~4.2). We omit this answer for brevity.

\section{Discussion: Classifying integrable $(2{+}1)$-dimensional systems of
hydrodynamic type}
\label{sec6}

The path from GT systems to integrable $(2{+}1)$-dimensional systems of
hydrodynamic type outlined in Secs.~\ref{sec3} and~\ref{sec4} can be
regarded as the following project of classifying integrable
$(2{+}1)$-dimensional systems.

\begin{step}
\label{step1}
Classify all GT systems~\eqref{28} (up to equivalence relations~\eqref{29}
and~\eqref{30}).
\end{step}

\begin{step}
\label{step2}
Classify all the nontrivial solutions $F,G$ of functional
equation~\eqref{25} for each GT system.
\end{step}

\begin{step}
\label{step3}
Use the lemma construction to construct the $(2{+}1)$-dimensional systems
corresponding to each GT system and each solution of Eq.~\eqref{25}.
\end{step}

In the following remarks, we share our experience acquired during several
years of working with integrable systems~\eqref{1}. The major part of these
notes are observations and conjectures. We currently have few rigorously
proved statements. Most of unproved statements are difficult. In particular,
the problem of classifying $n$-field GT systems with arbitrary $n$ is very
difficult. But such a classification is apparently possible for lower $n$.
We intend to study this problem in the future.

\medskip

{\bf Remarks concerning Step~\ref{step1}}

\smallskip

1. The different general-position GT systems are few and are universal:
several families of $(2{+}1)$-dimensional systems, each family depending on
essential parameters, correspond to each of the GT systems. The known GT
systems correspond to algebraic curves. Apparently, for any genus
$g=0,1,\dots$, we have a unique GT system with one field for $g=0,1$ and
$3g-3$ fields for $g>1$ (see Examples~\ref{ex2}--\ref{ex4}) such that every
integrable general-position $(2{+}1)$-dimensional system corresponds to
either this GT system or its regular extension. From the algebraic-geometric
standpoint, a regular extension corresponds to adding a marked point on the
curve.

\smallskip

2. The definition of GT system using formulas~\eqref{28} admits a
coordinate-free reformulation. Let $M$ be a fiber bundle with a
one-dimensional fiber $E$ and an $n$-dimensional base $F$. Every $p_i$ is
then a coordinate on $E$, and $u_1,\dots,u_n$ are the coordinates on $F$. We
can thus define the notion of the GT structure on $M$. We then apparently
have a canonical GT structure on the natural fiber bundle over the moduli
space $M_g$ of genus-$g$ curves. A fiber $E$ of this fiber bundle is the
curve corresponding to a point in $M_g$.

\smallskip

3. Integrable systems~\eqref{1} are defined up to arbitrary pointwise
transformations $\Bu\to \Phi(\Bu)$. The equivalence problem for
systems~\eqref{1} is nontrivial; in particular, finding a coordinate system
in which a $(2{+}1)$-dimensional system has the simplest form is often
difficult. One possible approach to this problem is based on the observation
that the simplest coordinates for a GT system, as a rule, are simultaneously
the simplest ones for the corresponding $(2{+}1)$-dimensional system.

\smallskip

4. ``Proper" GT system coordinates admit an algebraic-geometric
interpretation. For example, we can interpret each of the variables $p_i$
and $u$ in Example~\ref{ex2} as a coordinate on $\mathbb C P^1$ and each of
the variables $p_i$ in Example~\ref{ex3} as a coordinate on an elliptic
curve with $\tau$ being the modular parameter of this curve (i.e., a
coordinate on the moduli space of genus-one curves). The field functions
$u$, $v$, and $w$ in Example~\ref{ex4} are coordinates on the moduli space
of genus-two curves. We expect that in the case $g>2$, we have a GT system
whose $3g{-}3$ field functions are coordinates on the moduli space $M_g$ of
genus-$g$ curves. Finding good formulas here is hindered by the absence of a
good choice of coordinates on $M_g$. We can obtain an algebraic-geometric
description of these GT systems simultaneously with the description of
hydrodynamic reductions of the $(2{+}1)$-dimensional systems from
Krichever's paper~\cite{6}. As soon as we find GT systems related to curves
of genus $g>2$, describing solutions of the corresponding functional
equation~\eqref{25} would apparently result in constructing
$(2{+}1)$-dimensional systems depending on essential parameters. We must
then obtain the first examples of generalized hypergeometric functions
related to higher-genus curves. This program was realized
in~\cite{20},~\cite{28} for $g=0,1$. We note that the $(2{+}1)$-dimensional
systems in~\cite{6} do not contain essential parameters and correspond to the
simplest integer values of parameters in the hypergeometric series.

\smallskip

5. The problem of describing all the GT systems with a given number of
fields includes the problem of describing degenerations of a general
position system. Degenerations appearing when merging the field functions
$u_i$ (merging the marked points in the algebraic-geometric language) can be
described using Young diagrams. In Sec.~3.2, we considered the degeneration
of the GT system in Example~\ref{ex2} under which all the points merge into
a single point. Gibbons--Tsarev system~\eqref{61} corresponds to
degeneration of the type $u_i\to\lambda_i+\vep u_i$, $\vep\to0$.

\medskip

{\bf Remarks concerning Step~\ref{step2}}

\smallskip

6. Algebraic curves in the classification of integrable
$(2{+}1)$-dimensional systems appear in Step~\ref{step2}. In all the known
examples, solutions $F,G$ of functional equation~\eqref{25} are
algebraically dependent as functions of $p$. The polynomial relation between
them is called the dispersion curve (see~\eqref{22}). If we are interested
in solutions $F,G$ parameterizing a curve of a given genus for a given GT
system, then we assume that the dependence of $F,G$ on $p$ is known. For
example, in the case of a rational curve, the functions $F,G$ are rational
functions of $p$. Different choices of the degrees of $F,G$ result in
different families of $(2{+}1)$-dimensional systems. The functional equation
becomes a system of nonlinear differential equations for the ansatz
coefficients after the corresponding ansatz is substituted.

\smallskip

7. If we additionally assume that system~\eqref{1} admits a pseudopotential
representation, then we can find solutions $F,G$ using functional
equation~\eqref{76}, which is simpler than~\eqref{25} (see
Appendix~\ref{pr2}).

\smallskip

8. The class of integrable systems~\eqref{1} is closed under changes of
variables~\eqref{11}. The existence of this group leads to formulas~\eqref{50}
for solutions $F,G$ of functional equation~\eqref{25}. In these formulas, the
expression for $S(h,p)$ is linear in $h$, and $h_1$, $h_2$, and $h_3$ are
arbitrary elements of a vector space $\cH$ of dimension $M$. In the general
position case, $\cH$ is the solution space for an overdetermined system of
linear partial differential equations of hypergeometric type. We can
interpret formulas~\eqref{50} as a linearizing substitution for the nonlinear
system mentioned in item~5. The space $\cH$ may consist of elementary
functions in the case of reduction. For example, the space $\cH$ consists of
quasipolynomials in Sec.~\ref{sec5}.

\smallskip

9. In the case of curves of genus $g=0,1$, the dimension of $\cH$ is $n+1$,
where $n$ is the number of marked points. The set of times $t_i$,
$i=1,\dots,n+1$, in $\cH$ corresponds to the basis $h_i$, $i=1,\dots,n+1$.
For any triple $t_i,t_j,t_k$ of times, formula~\eqref{50} results in a
three-dimensional system of form~\eqref{1} with the independent variables
$t_i$, $t_j$, and $t_k$. All the systems corresponding to different triples
of times are mutually compatible.

\appendix
\def\theequation{\arabic{equation}}
\setcounter{equation}{66}

\section{Conservation laws of hydrodynamic type and integrable quasilinear
second-order equations}
\label{pr1}

A relation of the form
$$
a(\Bu)_t+b(\Bu)_y+c(\Bu)_x=0,
$$
where $a$, $b$, and $c$ are some scalar functions, that is satisfied by any
solution $\Bu$ of system~\eqref{55} is called a conservation law of
hydrodynamic type for system~\eqref{1}. All known integrable
systems~\eqref{1} have a rich collection of conservation laws. More
precisely, we consider systems~\eqref{1} with $k=0$. It was proved
in~\cite{20} that systems~\eqref{55} corresponding to a rational nonsingular
curve (Example~\ref{ex1}) have $n{+}1$ linearly independent conservation
laws; all partial derivatives of the functions $a$, $b$, and $c$ with
respect to the $u_i$ were found explicitly there. The dimension of
the space of conservation laws is apparently $n$ in the case of an elliptic
curve (Example~\ref{ex3}). This was verified for degenerations of the
elliptic systems presented in~\cite{7}.

In the case of exponential systems~\eqref{64} with $k=0$, we can find
conservation laws explicitly.

\begin{proposition}
\label{prop4}
System~\eqref{64} with $k=0$ admits the $n{+}1$ conservation laws
$$
\begin{vmatrix}S^\rg_n(g_2,\lambda_i)&S^\rg_n(g_3,\lambda_i)\\[1mm]
a_{2,i}e^{-u_i}&a_{3,i}e^{-u_i}\end{vmatrix}_t+\begin{vmatrix}
S^\rg_n(g_3,\lambda_i)&S^\rg_n(g_1,\lambda_i)\\[1mm]
a_{3,i}e^{-u_i}&a_{1,i}e^{-u_i}\end{vmatrix}_y+\begin{vmatrix}
S^\rg_n(g_1,\lambda_i)&S^\rg_n(g_2,\lambda_i)\\[1mm]
a_{1,i}e^{-u_i}&a_{2,i}e^{-u_i}\end{vmatrix}_x=0,
$$
where $i=1,\dots,n$, and
$$
\begin{vmatrix}S^\rg_n(g_2,\lambda_0)&S^\rg_n(g_3,\lambda_0)\\[1mm]
a_{2,0}&a_{3,0}\end{vmatrix}_t+\begin{vmatrix}
S^\rg_n(g_3,\lambda_0)&S^\rg_n(g_1,\lambda_0)\\[1mm]
a_{3,0}&a_{1,0}\end{vmatrix}_y+\begin{vmatrix}
S^\rg_n(g_1,\lambda_0)&S^\rg_n(g_2,\lambda_0)\\[1mm]
a_{1,0}&a_{2,0}\end{vmatrix}_x=0.
$$
Here,
$$
S^\rg_n(g,\lambda_i)=\biggl(S_n(g,p)-\frac{a_ie^{u_i}}
{p-\lambda_i}\biggr)\bigg|_{p=\lambda_i},\qquad
S^\rg_n(g,\lambda_0)=\biggl(S_n(g,p)-
\frac{a_0}{p-\lambda_0}\biggr)\bigg|_{p=\lambda_0}.
$$
\end{proposition}

In the case $k>0$, all known integrable systems~\eqref{1} have $k$ triples
of ``short" conservation laws of the form
$$
a_y=b_x,\qquad a_t=c_x,\qquad b_t=c_y.
$$
We can introduce the potential $Z_x=a$, $Z_y=b$, and $Z_t=c$ for each such
triple. For small $k$ (for $n\ge2k$ in the rational case and for $n\ge2k-1$
in the elliptic case), introducing potentials $Z_1,\dots,Z_k$, we make
system~\eqref{1} a properly defined system consisting of several
second-order equations and several first-order equations. In particular, for
$n=3k$, system~\eqref{1} is equivalent to a system of the form
\begin{equation}
A_1\BZ_{tt}+A_2\BZ_{yt}+A_3\BZ_{xt}+A_4\BZ_{yy}+A_5\BZ_{xy}+A_6\BZ_{xx}=0,
\label{67}
\end{equation}
where $\BZ=(Z_1,\dots,Z_k)^{\mathrm T}$ and $A_i$ are some $k{\times}k$
matrices depending on $\BZ_x$, $\BZ_y$, and $\BZ_t$.

Short conservation laws for the systems in Sec.~4.1 with $k>0$ are given by
the formulas
\begin{equation}
\biggl(\frac{\Delta(g_r,h_1,\dots,\hat{i},\dots,h_k)}
{\Delta(h_1,\dots,h_k)}\biggr)_{t_s}=
\biggl(\frac{\Delta(g_s,h_1,\dots,\hat{i},\dots,h_k)}
{\Delta(h_1,\dots,h_k)}\biggr)_{t_r},
\label{68}
\end{equation}
where $i=1,\dots,k$, $r,s=1,2,3$, $t_1=t$, $t_2=y$, $t_3=x$, and
$$
\Delta(f_1,\dots,f_k)=\det\begin{pmatrix}f_1&f_2&\cdots&f_k\\[1mm]
f_{1,u_{n-k+2}}&f_{2,u_{n-k+2}}&\cdots&f_{k,u_{n-k+2}}\\[1mm]
\vdots&\vdots&\ddots&\vdots\\[1mm]
f_{1,u_n}&f_{2,u_n}&\cdots&f_{k,u_n}\end{pmatrix}.
$$
In addition to these conservation laws, such systems apparently also have
$n{+}1$ linearly independent conservation laws of hydrodynamic type.

In the case where $n=3$ and $k=1$, the system in Sec.~4.1 depends on four
arbitrary linearly independent constant vectors (see~\eqref{51}). After the
potential is introduced using the short conservation laws, the system
becomes an equation of form~\eqref{2} whose coefficients $A_i$ are
polynomials of a degree not exceeding two in the variables $Z_x$, $Z_y$,
and $Z_t$. Using the group $GL(4)$ acting on the set of equations~\eqref{2},
we can reduce the matrix composed from the above vectors to the unit matrix
by linear transformations of the variables $Z$, $x$, $t$, and $y$. The most
convenient way to do this is to make the function $h$ equal to unity. Then,
in the general position case where $n_0=0$ and $n_1=n_2=n_3=1$
(see~\eqref{65}), the equation coincides with~\eqref{7}. The degeneration of
the form $n_0=3$ and $n_1=n_2=n_3=0$ results in the equation~\cite{41}
$$
Z_{xt}-Z_{yy}+Z_x Z_{xy}-Z_yZ_{xx}=0,
$$
the degeneration $n_0=2$, $n_1=1$, and $n_2=n_3=0$ gives
$$
Z_{yt}+Z_t Z_{xx}-Z_x Z_{xt}=0,
$$
the degeneration $n_0=1$, $n_2=2$, and $n_2=n_3=0$ gives
$$
Z_y Z_{xt}-Z_t Z_{xy}+Z_{yy}=0,
$$
and the degeneration $n_0=n_1=n_2=1$ and $n_3=0$ gives
$$
Z_y Z_{xt}-Z_t Z_{xy}+Z_{yt}=0.
$$

\section{The GT systems and the pseudopotential representation}
\label{pr2}

In Sec.~\ref{sec4}, we described several classes of solutions of functional
equation~\eqref{25} for the rational and elliptic GT systems. These
solutions were found in~\cite{20},~\cite{28} under the additional assumption
that in addition to hydrodynamic reductions, the corresponding
$(2{+}1)$-dimensional systems~\eqref{1} have the pseudopotential
representation or, equivalently, the dispersionless Lax representation. As
we demonstrate below, this assumption results in a functional equation that
is simpler than~\eqref{25} and was in fact solved.

The dispersionless Lax representation is the relation
\begin{equation}
L_t=\{L ,A\},
\label{69}
\end{equation}
where $\{L,A\}=A_{\lambda}L_x-A_xL_{\lambda}$. Here, $A=A(\lambda,u_1(x,t),
\dots,u_n(x,t))$ and the unknown function $L=L(\lambda,u_1(x,t),\dots,
u_n(x,t))$ depend on the spectral parameter $\lambda$. The transformation
$L(x,t,\lambda)\to\lambda(x,t,L)$ takes~\eqref{69} to the conservative form
\begin{equation}
\lambda_t=A(\lambda,u_1,\dots,u_n)_x,
\label{70}
\end{equation}
where $L$ now plays the role of a hidden parameter. We can rewrite the
last equation as
\begin{equation}
\Phi_t=A(\Phi_x,u_1,\dots,u_n),
\label{71}
\end{equation}
where $\lambda=\Phi_x$.

We assume that system~\eqref{1} has both hydrodynamic reductions~\eqref{24}
and pseudopotential representation~\eqref{4}. Calculating $L_t$ and $L_x$ on
the $N$-phase solutions and using~\eqref{43}, we write~\eqref{69} as
$$
\sum_iF(p_i,u_1,\dots,u_n)\ptl_iLr^i_x=\sum_i\biggl(
\ptl_iL A_{\lambda}(\lambda,u_1,\dots,u_n)-L_{\lambda}
\sum_{k=1}^nA_{u_k}(\lambda,u_1,\dots,u_n)\ptl_iu_k\!\biggr)r^i_x.
$$
Equating the coefficients of $r^i_x$, we obtain
$$
F(p_i,u_1,\dots,u_n)\ptl_i L=A_{\lambda}(\lambda,u_1,\dots,u_n)\ptl_i L-
L_{\lambda}\sum_{k=1}^nA_{u_k}(\lambda,u_1,\dots,u_n)\ptl_iu_k.
$$
The function $L$ therefore satisfies the system of equations
\begin{equation}
\ptl_iL=t(p_i,\lambda,u_1,\dots,u_n)L_{\lambda}\ptl_iu_1,\quad
i=1,\dots,N,
\label{72}
\end{equation}
which is compatible by virtue of the corresponding GT system~\eqref{28}.
System~\eqref{72} is called the {\sl L\"owner equation.}

Using transformation~\eqref{29}, we can arbitrarily gauge the function $F$.
For example, we can set $F=p_i$. It is convenient to choose
$$
F(p_i,u_1,\dots,u_n)=A_{p_i}(p_i,u_1,\dots,u_n),
$$
and system~\eqref{72} then becomes
\begin{equation}
\ptl_iL=\frac{\sum_{k=1}^nA_{u_k}(\lambda,u_1,\dots,u_n)\ptl_iu_k}
{A_{\lambda}(\lambda,u_1,\dots,u_n)-A_{p_i}(p_i,u_1,\dots,u_n)}L_{\lambda},
\quad i=1,\dots,N.
\label{73}
\end{equation}
The transformation $L(x,t,\lambda)\to \lambda(x,t,L)$ brings~\eqref{73} to
the form
\begin{equation}
\ptl_i \lambda =\frac{\sum_{k=1}^nA_{u_k}(\lambda,u_1,\dots,u_n)\ptl_iu_k}
{A_{p_i}(p_i,u_1,\dots,u_n)-A_{\lambda}(\lambda,u_1,\dots,u_n)},\quad
i=1,\dots,N.
\label{74}
\end{equation}

Relation~\eqref{73} implies that $L_{\lambda}=0$ at the points $\lambda=
p_i$, $i=1,\dots,N$. Substituting $\lambda=p_i$ in~\eqref{69}, we obtain
$$
(L(p_i,r^1,\dots,r^N))_t=A_{p_i}(p_i,u_1,\dots,u_n)(L(p_i,r^1,\dots,r^N))_x.
$$
We see that the functions $L(p_i,r^1,\dots,r^N)$ and $r_i$ satisfy
system~\eqref{43} with the same functions $F_i$. Hence, $L(p_i,r^1,\dots,
r^N)=\mu_i(r^i)$ for some functions $\mu_i$. After changing the Riemann
invariants $\mu_i(r^i)\to r^i$, we can set
$$
r^i=L(p_i,r^1,\dots,r^N).
$$
After the transformation $L(x,t,\lambda)\to\lambda(x,t,L)$, this relation
becomes
$$
p_i=\lambda(x,t,r^i).
$$
Substituting $L=p_j$ in~\eqref{74}, we obtain the relation
\begin{equation}
\ptl_ip_j=\frac{\sum_{k=1}^nA_{u_k}(p_j,u_1,\dots,u_n)\ptl_iu_k}
{A_{p_i}(p_i,u_1,\dots,u_n)-A_{p_j}(p_j,u_1,\dots,u_n)},
\quad i\ne j,\quad i,j=1,\dots,N,
\label{75}
\end{equation}
which expresses the function $f(p_i,p_j,u_1,\dots,u_n)$ in the corresponding
GT system~\eqref{28} in terms of the potential $A$ and the functions
$g_2,\dots,g_n$. This formula for $n=1$ is contained in~\cite{42}.

We recall that the GT-system gauge is fixed by the condition $F=A_{p_i}$ in
formula~\eqref{75}. This gauge is generally inconvenient. After an arbitrary
transformation~\eqref{29}, formula~\eqref{75} becomes
\begin{equation}
\ptl_ip_j=\frac{\sum_{k=1}^n\bigl(\Phi_{p_i}(p_i,u_1,\dots,u_n)\phi_{u_k}
(p_j,u_1,\dots,u_n)-\phi_{p_i}(p_i,u_1,\dots,u_n)
\Phi_{u_k}(p_j,u_1,\dots,u_n)\bigr)\ptl_iu_k}
{\phi_{p_i}(p_i,u_1,\dots,u_n)\Phi_{p_j}(p_j,u_1,\dots,u_n)-
\phi_{p_j}(p_j,u_1,\dots,u_n)\Phi_{p_i}(p_i,u_1,\dots,u_n)},
\label{76}
\end{equation}
where $i\ne j$, $i,j=1,\dots,N$, and the potential is defined parametrically
$$
A=\phi(p,u_1,\dots,u_n),\qquad\lambda=\Phi(p,u_1,\dots,u_n).
$$

Using~\eqref{28} and omitting the arguments $u_1,\dots,u_n$ for brevity, we
rewrite formula~\eqref{76} in the form
\begin{equation}
f(p_i,p_j)=\frac{\sum_{k=1}^n(\Phi_{p_i}(p_i)\phi_{u_k}(p_j)-
\phi_{p_i}(p_i)\Phi_{u_k}(p_j))g_k(p_i)}
{\phi_{p_i}(p_i)\Phi_{p_j}(p_j)-\phi_{p_j}(p_j)\Phi_{p_i}(p_i)},
\label{77}
\end{equation}
where $g_1=1$. For a fixed GT system~\eqref{28}, this relation is a
functional equation for the functions $\phi$ and $\Phi$ that is much simpler
than functional equation~\eqref{25} for the functions $F$ and $G$. In
particular, fixing $p_j$, we can find the dependence of the ratio
$\phi_{p_i}/\Phi_{p_i}$ on the variable $p_i$ from~\eqref{77}. Every
solution of Eq.~\eqref{77} determines an {\sl integrable potential
corresponding to system}~\eqref{28}.

\medskip

{\bf Example 2a} (continuation of Example~\ref{2}){\bf.}
We find the one-field integrable potentials corresponding to one-field GT
system~\eqref{34} with $P(x)=x(x-1)$. Fixing $p_j$ in~\eqref{77}, we find
that $\phi_{p_i}/\Phi_{p_i}$ is the ratio of polynomials of the first degree
in $p_i$. We set
$$
\Phi_{p_i}=(A_1(u)p_i+A_0(u))Z(p_i,u),\qquad
\phi_{p_i}=(B_1(u)p_i+B_0(u))Z(p_i,u).
$$
Substituting these expressions in~\eqref{77} and equating the coefficients
of like powers of $p_j$, we find that~\eqref{77} is equivalent to the
relations
\begin{align*}
&\Phi_u=-(A_1(u)u+A_0(u))\frac{p_i(p_i-1)}{u(u-1)}\,Z(p_i,u),
\\[2mm]
&\phi_u=-(B_1(u)u+B_0(u))\frac{p_i(p_i-1)}{u(u-1)}\,Z(p_i,u).
\end{align*}
We find both partial derivatives of the function $Z(p_i,u)$ from the
compatibility condition for these two systems (i.e., from the condition that
the mixed derivatives of $\phi$ and $\Phi$ are equal). We then find the
condition for the equality of the mixed derivatives of $Z$. Equating the
coefficients of like powers of $p_i$ in this relation, we obtain a system of
nonlinear ODEs for the functions $A_i$ and $B_i$, $i=0,1$. We can express a
solution of this system in terms of two arbitrary solutions $y_1$ and $y_2$
of the standard hypergeometric equation
$$
u(u-1)y(u)''+ [(\alpha+\beta+1)u-\gamma]y(u)'+\alpha\beta y(u)=0
$$
as
$$
A_1=-\alpha y_1,\qquad A_0=u(u-1)y_1'+\alpha uy_1,\qquad
B_1=-\alpha y_2,\qquad B_0=u(u-1)y_2'+\alpha uy_2.
$$
We reconstruct the function $Z$ from its partial derivatives:
$$
Z(p,u)=(p-1)^{\alpha+\gamma}p^{-\beta-\gamma-1}(p-u)^{\beta-1}.
$$
We finally obtain
\begin{equation}
\Phi(p,u)=\int_0^p[u(u-1)y_1'(u)+\alpha(u-\xi)y_1(u)]
(\xi-1)^{\alpha+\gamma}\xi^{-\beta-\gamma-1}(\xi-u)^{\beta-1}\,d\xi.
\label{78}
\end{equation}
To obtain the function $\phi$, we merely substitute $y_2$ for $y_1$ in this
formula.

\medskip

The potential found above corresponds to the general position case
in~\cite{27}, where all the integrable one-field potentials were described.
The generalization of formula~\eqref{71} to the case of $n$-field GT
system~\eqref{45},~\eqref{46} is
$$
P_n(p,h)=\int_0^pS_n(h,\xi)(\xi-u_1)^{-s_1-1}\cdots(\xi-u_n)^{-s_n-1}
\xi^{-s_{n+1}-1}(\xi-1)^{-s_{n+2}-1}\,d\xi,
$$
where $h(u_1,\dots,u_n)$ is an arbitrary solution of
system~\eqref{47},~\eqref{48} and the polynomial $S_n$ is given by
formula~\eqref{49}. The elliptic analogue of this potential is given by
$$
P_n(p,h)=\int_0^pS_n(h,\xi)e^{2\pi ir(\tau-\xi)}
\frac{\theta'(0)^{-s_1-\dots-s_n}\theta(u_1)^{s_1}\cdots\theta(u_n)^{s_n}}
{\theta(\xi)^{-s_1-\dotsb-s_n}\theta(\xi-u_1)^{s_1}\cdots
\theta(\xi-u_n)^{s_n}}\,d\xi,
$$
where we use the notation in Example~\ref{ex3}. Integrable potentials with
the defect index $k>0$ corresponding to rational and elliptic curves were
presented in~\cite{20},~\cite{28}.

Clearly, for any solution $\phi,\Phi$ of Eq.~\eqref{77}, the pair
$k_1\phi+k_2\Phi,k_3\phi+k_4\Phi$ is a solution for any constant $k_i$. In
the above example, such a transformation corresponds to another choice of
the hypergeometric equation solutions $y_1,y_2$. The function $\phi$
therefore always lies in a two-dimensional space $V$. If this space has the
dimension $d\ge3$, then choosing linearly independent elements $\phi_1,
\phi_2,\phi_3\in V$, we obtain the functions
$$
F=\frac{(\phi_1)_{p_i}}{(\phi_3)_{p_i}},\qquad
G=\frac{(\phi_2)_{p_i}}{(\phi_3)_{p_i}}
$$
satisfying functional equation~\eqref{25}. This is how solutions of
Eq.~\eqref{25} were found in~\cite{20},~\cite{28}. Equation~\eqref{25} for
GT system~\eqref{61} was solved directly.

\section{Multidimensional integrable systems of hydrodynamic type}
\label{pr3}

Multidimensional generalizations of systems~\eqref{1} are systems of the
form
\begin{equation}
\sum_{i=1}^dA_i(\Bu)\frac{\ptl \Bu}{\ptl x_i}=0,\quad d>3,
\label{79}
\end{equation}
where $A_i$ are $n{\times}n$ matrices and $\Bu=(u_1,\dots,u_n)$. We do not
here discuss the definition of the integrability of such systems based on
the hydrodynamic reduction method (see~\cite{15}). But if we know what
integrable system~\eqref{1} is, we can naturally require that every
reduction
\begin{equation}
\Bu=\Bu\biggl(\sum_{i=1}^dk_{1,i}x_i,\;
\sum_{i=1}^d k_{2,i}x_i,\;\sum_{i=1}^dk_{3,i}x_i\biggr)
\label{80}
\end{equation}
of integrable system~\eqref{79} result in a three-dimensional integrable
system~\eqref{1} for any constants $k_{i,j}$. We note that the coefficients
of this system of form~\eqref{1} depend on the parameters $k_{i,j}$ linearly.
It was shown in~\cite{43} that in the case $n=2$, systems~\eqref{79} have
this property, being reductions of the system
\begin{equation}
u_{x_1}+v_{x_2}+uv_{x_3}-vu_{x_3}=0,\qquad
u_{x_4}+v_{x_5}+uv_{x_6}-vu_{x_6}=0.
\label{81}
\end{equation}

System~\eqref{81} admits a maximum collection of hydrodynamic reductions and
can be written as the commutation condition for a pair of vector fields.
This system was first derived in~\cite{15} as follows. We have four
parameters (namely, the coefficients $p_0$, $p_1$, $p_2$, and $p_3$ of the
polynomial $P$) in system~\eqref{58} entering the system coefficients
linearly. If we write~\eqref{58} as
$$
\Bu_t+A(\Bu)\Bu_x+\sum_{i=0}^3p_iB_i(\Bu)\Bu_y=0
$$
and pass to the six-dimensional system
$$
\Bu_t+A(\Bu)\Bu_x+\sum_{i=0}^3B_i(\Bu)\Bu_{y_i}=0,
$$
then this last system is equivalent to system~\eqref{81}.

Analyzing system~\eqref{58} and other known examples of multidimensional
integrable dispersionless systems, we proposed the conjecture that {\sl for
any system~\eqref{1} obtained by reduction~\eqref{80} from an integrable
multidimensional system, the corresponding GT system is equivalent to
system~\eqref{37}}~\cite{44}.

In Sec.~\ref{sec5}, we constructed a broad class of systems~\eqref{1} whose
corresponding GT systems have form~\eqref{37}. Systems~\eqref{64} are the
simplest ones. Below, we attempt to construct multidimensional integrable
systems corresponding to these systems.

System~\eqref{64} is written in the coordinates $u_i$ in which the
corresponding GT system acquires the simplest form. According to
Proposition~\ref{prop3}, system~\eqref{64} admits a pseudopotential
representation depending rationally on the spectral parameter. Taking zeros
of the pseudopotential representation denominator as the new field functions
$v_i$, $i=1,\dots,n$, we reduce the system to the form analogous
to~\eqref{58}
$$
(v_i)_t-\frac{f_{v_i}}{g_{v_i}}(v_i)_x+\biggl(\sum_{k\ne i}\gamma_{ik}
P(v_k)+\delta_i\biggr)(v_i)_y+\sum_{j\ne i}\beta_{ij}P(v_i)(v_j)_y=0,\quad
i=1,\dots,n,
$$
where $\beta_{ij}$, $\gamma_{ik}$, and $\delta_i$ are expressed in terms of
the two symmetric polynomials
\begin{align*}
&f(\Bv)=f_0+f_1(v_1+\dots+v_n)+\dots+f_nv_1\cdots v_n,
\\[2mm]
&g(\Bv)=g_0+g_1(v_1+\dots+v_n)+\dots+g_nv_1\cdots v_n
\end{align*}
(which are linear in each of the variables) and the polynomial
$P(x)=x^{n+1}+p_n x^n+\dots+p_0$ as
\begin{align*}
&\beta_{ij}=\frac{f_{v_iv_j}g_{v_i}-g_{v_iv_j}f_{v_i}}
{(v_i-v_1)\cdots(v_i-v_n)g_{v_i}},\qquad
\gamma_{ik}=\frac{f_{v_i}g_{v_k}-g_{v_i}f_{v_k}}
{(v_i-v_1)\cdots(v_i-v_n)g_{v_i}(v_i-v_k)},
\\[2mm]
&\delta_i=\frac{g f_{v_i}-f g_{v_i}}{g_{v_i}}.
\end{align*}

\begin{gyph}
\label{gyp1}
The corresponding $(n{+}4)$-dimensional system
\begin{equation}
(v_i)_t-\frac{f_{v_i}}{g_{v_i}}(v_i)_x+\sum^{n+1}_{\substack{m=0,\\k\ne i}}
\gamma_{ik}v_k^m(v_i)_{y_m}+\delta_i (v_i)_{y_{n+1}}+
\sum^{n+1}_{\substack{m=0,\\j\ne i}}\beta_{ij}v_i^m(v_j)_{y_m}=0,
\label{82}
\end{equation}
$i=1,\dots,n$, is integrable for any $n$.
\end{gyph}

We note that for $n=3$, we can easily reduce system~\eqref{82} to the form
analogous to~\eqref{81}
\begin{align*}
&u_{x_1}-w_{x_2}+uw_{x_3}-wu_{x_3}=0,\qquad
v_{y_1}-w_{y_2}+vw_{y_3}-wv_{y_3}=0,
\\[2mm]
&w_z=v_{x_1}-u_{y_1}+vw_{x_3}-wv_{x_3}-uw_{y_3}+wu_{y_3}.
\end{align*}

We plan to reduce system~\eqref{82} to a simpler form for arbitrary $n$, to
prove that it is integrable, and to study the corresponding GT system in a
separate paper. Here, we only note that the definition of the GT system as a
compatible system of form~\eqref{28} can be easily generalized to the case
$d>3$. Moreover, if GT systems in the three-dimensional case are related to
fiber bundles with a one-dimensional fiber (see Sec.~\ref{sec6}), then a
fiber in the $d$-dimensional case has the dimension $d-2$. Some GT systems
for $d>3$ were found in~\cite{15}. We present a few new examples of
one-field GT systems in the case $d=4$.

Let $P(x)=z_2x^2+z_1x+z_0$ and $Q(x)$ be two arbitrary quadratic
polynomials, $J(x)$ be an arbitrary linear polynomial, and $S(x,y)=2z_2xy+
z_1(x+y)+2z_0$.

\begin{example}
\label{ex8}
The formulas
\begin{align*}
&\ptl_ip_j=\frac{(p_i-p_j)^2}{P(p_i)(q_i-q_j)}\ptl_iu,\qquad
\ptl_iq_j=\frac{Q(p_i)}{P(p_i)}\ptl_i u,
\\[2mm]
&\ptl_i\ptl_ju=\frac{(p_i-p_j)S(p_i,p_j)}
{P(p_i)P(p_j)(q_i-q_j)}\ptl_iu\ptl_ju
\end{align*}
give a one-field GT system with a two-dimensional fiber. The coordinates in
the fiber are $p_i$ and $q_i$, and $u$ is the coordinate on the base. Using a
fractional-linear transformation of $p$, we can reduce the polynomial $P$ to
one of the two canonical forms $P(x)=x$ or $P(x)=1$. The case $P(x)=1$ in
other coordinates arose in~\cite{15} when investigating the equation
$$
Z_{tx}+Z_{xy}+Z_{xx}Z_{yy}-Z_{xy}^2=0.
$$
The general position case $P(x)=x$ is possibly related to the most
nondegenerate integrable equation of this sort.
\end{example}

\begin{example}
\label{ex9}
The formulas
\begin{align*}
&\ptl_ip_j=\frac{(p_i-p_j)J(p_i)J(p_j)}{P(p_i)(q_i-q_j)^2}\ptl_iu,\qquad
\ptl_iq_j=\frac{J(p_i)^2}{P(p_i)(q_i-q_j)}\ptl_iu,
\\[2mm]
&\ptl_i\ptl_ju=\frac{J(p_i)J(p_j)S(p_i,p_j)}{P(p_i)P(p_j)(q_i-q_j)^2}
\ptl_iu\ptl_ju
\end{align*}
give a one-field GT system.
\end{example}

\begin{example}
\label{ex10}
The formulas
\begin{align*}
&\ptl_ip_j=\frac{p_i p_j(2-p_j)}{(q_i-q_j)^2}\ptl_iu,\qquad
\ptl_iq_j=\frac{p_i}{q_i-q_j}\ptl_iu,
\\[2mm]
&\ptl_i\ptl_ju=\frac{p_i p_j}{(q_i-q_j)^2}\ptl_iu\ptl_ju
\end{align*}
give a one-field GT system.
\end{example}

We do not know whether the GT systems in Examples~\ref{ex9} and~\ref{ex10}
are related to some multidimensional integrable dispersionless systems.

\section{Classifying integrable chains of hydrodynamic type}
\label{pr4}

We consider integrable quasilinear infinite chains of the form
\begin{equation}
u_{\alpha,t}=\phi_{\alpha,1}u_{1,x}+\dots+
\phi_{\alpha,\alpha+1}u_{\alpha+1,x},\quad
\alpha=1,2,\dots,\quad\phi_{\alpha,\alpha+1}\ne0,
\label{83}
\end{equation}
where $\phi_{\alpha,j}=\phi_{\alpha,j}(u_1,\dots,u_{\alpha+1})$. Two chains
are {\sl equivalent} if they are related by a transformation of the form
\begin{equation}
u_{\alpha}\to\Psi_{\alpha}(u_1,\dots,u_{\alpha}),\quad
\frac{\ptl\Psi_{\alpha}}{\ptl u_{\alpha}}\ne0,\quad\alpha=1,2,\ldots.
\label{84}
\end{equation}
We consider a chain integrable if it admits hydrodynamic reductions
(see~\cite{13}--\cite{15},~\cite{23}--\cite{25}).

\begin{example}
\label{ex11}
The Benney chain~\cite{45}--\cite{47}
\begin{equation}
u_{1,t}=u_{2,x},\qquad
u_{2,t}=u_1u_{1,x}+u_{3,x},\qquad\dots,\qquad
u_{\alpha t}=(\alpha-1)u_{\alpha-1}u_{1,x}+u_{\alpha+1,x},\qquad\dots
\label{85}
\end{equation}
is the best-known example of an integrable chain~\eqref{83}. Hydrodynamic
reductions of the Benney chain were studied in~\cite{12}.
\end{example}

\begin{definition}
\label{def9}
A {\sl hydrodynamic $(1{+}1)$-dimensional $N$-component reduction of
chain}~\eqref{83} is semi-Hamiltonian system~\eqref{43} in which the
functions $u_j(r^1,\dots,r^N)$, $j=1,2,\dots$, satisfy system~\eqref{83} for
every solution~\eqref{43}.
\end{definition}

Using transformations~\eqref{29}, we can set
\begin{equation}
F(p,u_1,\dots,u_n)=p
\label{86}
\end{equation}
without restricting the generality.

A calculation analogous to the one at the end of Sec.~\ref{sec2} results in
an infinite triangular GT system corresponding to the given integrable
chain.

\begin{definition}
\label{def10}
We call a compatible system of the form
\begin{equation}
\begin{aligned}
&\ptl_ip_j=f(p_i,p_j,u_1,\dots,u_n)\ptl_iu_1,
\\[2mm]
&\ptl_iu_k=g_k(p_i,u_1,\dots,u_k)\ptl_i u_1,\quad k=1,2,\dots,
\\[2mm]
&\ptl_i\ptl_ju_1=h(p_i,p_j,u_1,\dots,u_n)\ptl_iu_1\ptl_ju_1,
\end{aligned}
\label{87}
\end{equation}
where $i,j=1,\dots,N$, $i\ne j$, a {\sl triangular GT system}. Here,
$p_1,\dots,p_N$ and $u_1,u_2,\dots$ are functions of $r^1,\dots,r^N$ and
$\ptl_i=\ptl/\ptl r^i$.
\end{definition}

Substituting $u_i=u_i(r^1,\dots,r^N)$, $i=1,2,\dots$, in the chain,
calculating the derivatives with respect to $t$ and $x$ by virtue
of~\eqref{43}, and equating the coefficients of $r^s_x$ to zero, we obtain
$$
\ptl_su_{\alpha}p_s=\phi_{\alpha,1}\ptl_su_1+\dots+
\phi_{\alpha,\alpha+1}\ptl_su_{\alpha+1},\quad\alpha=1,2,\ldots.
$$
Using~\eqref{87} and replacing $p_s$ with $p$, we find
$$
p=\phi_{1,1}+\phi_{1,2}g_2,\qquad
pg_2=\phi_{2,1}+\phi_{2,2}g_2+\phi_{2,3}g_3,\qquad
pg_3=\phi_{3,1}+\phi_{3,2}g_2+\phi_{3,3}g_3+\phi_{3,4}g_4,\qquad\ldots.
$$
Solving this system for $g_2,g_3,\dots$, we obtain
$$
g_i(p)=\psi_{i,0}+\psi_{i,1}p+\dots+\psi_{i,i-1}p^{i-1},
$$
where $\psi_{i,j}$ are some functions of $u_1,\dots,u_i$. In particular,
\begin{equation}
g_2=-\frac p{\phi_{1,2}}-\frac{\phi_{1,1}}{\phi_{1,2}}.
\label{88}
\end{equation}

\medskip

{\bf Example 11a} (continuation of Example~\ref{ex11}){\bf.}
The triangular GT system corresponding to the Benney chain is
\begin{align}
&\ptl_ip_j=\frac{\ptl_iu_1}{p_i-p_j},\qquad
\ptl_i\ptl_ju_1=\frac{2\ptl_iu_1\ptl_ju_1}{(p_i-p_j)^2},
\label{89}
\\[2mm]
&\ptl_iu_m=\bigl(-(m-2)u_{m-2}-\dots-2u_2p_i^{m-2}-
u_1p_i^{m-3}+p_i^{m-1}\bigr)\ptl_iu_1.
\label{90}
\end{align}
Equations~\eqref{89} were first obtained in~\cite{12}.

\medskip

The compatibility conditions $\ptl_i\ptl_ju_{\alpha}=
\ptl_j\ptl_iu_{\alpha}$, $\alpha=2,3,4$, result in a system of linear
equations for $\ptl_ip_j$, $\ptl_jp_i$, and $\ptl_i\,\ptl_ju_1$. Solving it,
we obtain
\begin{align}
&\ptl_ip_j=\frac{P(p_i,p_j)}{p_i-p_j}\ptl_iu_1,\quad i\ne j,
\label{91}
\\[2mm]
&\ptl_i\ptl_ju_1=\frac{Q(p_i,p_j)}{(p_i-p_j)^2}\ptl_iu_1\ptl_ju_1,\quad
i\ne j,
\label{92}
\end{align}
where $P$ and $Q$ are quadratic polynomials in each of the variables $p_i$
and $p_j$. Their coefficients may in principle depend on $u_1$, $u_2$,
$u_3$, and $u_4$. But it is easy to deduce from the compatibility conditions
$\ptl_i\ptl_jp_k=\ptl_j\ptl_ip_k$ that $P$ and $Q$ depend only on $u_1$ and
$u_2$.

We write~\eqref{91} in the form
\begin{equation}
\ptl_ip_j=\biggl(\frac{R(p_j)}{p_i-p_j}+(z_4p_j^2+z_5p_j+z_6)p_i+
z_4p_j^3+z_3p_j^2+z_7p_j+z_8\biggr)\ptl_iu_1,
\label{93}
\end{equation}
where $R(x)=z_4x^4+z_3x^3+z_2x^2+z_1x+z_0$. From the GT-system
compatibility conditions, we easily obtain
\begin{equation}
\ptl_i\ptl_ju_1=\biggl(\frac{2z_4p_i^2p_j^2+z_3p_ip_j(p_i+p_j)+z_2
(p_i^2+p_j^2)+z_1(p_i+p_j)+2z_0}{(p_i-p_j)^2}+z_9\biggr)\ptl_iu_1\ptl_ju_1.
\label{94}
\end{equation}
Using change of variables~\eqref{84}, we can set the coefficient $z_9$ equal
to $z_6-z_7$. The coefficients $z_i(x,y)$, $i=1,\dots,8$, satisfy a pair of
compatible dynamical systems in $y$ and $x$. The first of these systems is
\begin{alignat*}2
&z_{0,y}=2z_0z_5-z_1z_6,&\qquad&z_{1,y}=4z_0z_4+z_1z_5-2z_2z_6,
\\[2mm]
&z_{2,y}=3z_1z_4-3z_3z_6,&\qquad&z_{3,y}=2z_2z_4-z_3z_5-4z_4z_6,
\\[2mm]
&z_{4,y}=z_3z_4-2z_4z_5,&\qquad&z_{5,y}=z_4z_7-z_4z_6-z_5^2,
\\[2mm]
&z_{6,y}=z_4z_8-z_5z_6,&\qquad&z_{7,y}=2z_1z_4-2z_3z_6-z_5z_6+z_4z_8,
\\[2mm]
&z_{8,y}=2z_0z_4-z_6^2-z_6z_7+z_5z_8.
\end{alignat*}
The second of these systems looks more complicated. To linearize these
systems, we reduce the polynomial $R$ to a canonical form, sacrificing
normalization~\eqref{86} for this.

It turns out that if the transformation coefficients
\begin{equation}
p_i=\frac{a\bar p_i+b}{\bar p_i-\psi},\quad i=1,\dots,N,
\label{95}
\end{equation}
satisfy the conditions
$$
a_{u_2}=z_4(b+a \psi),\qquad
b_{u_2}=z_4b\psi+z_5b-z_6a,\qquad\psi_{u_2}=z_4\psi^2+z_5\psi+z_6,
$$
then the transformation preserves the form of Eqs.~\eqref{93}
and~\eqref{94}. The polynomial $R$ changes simply under
transformations~\eqref{95}:
$$
R(p_i)\to(p_i-\psi)^4R\biggl(\frac{ap_i+b}{p_i-\psi}\biggr).
$$
We first assume that all the roots of $R$ are distinct. We can then verify
that using admissible transformations~\eqref{95}, we can set three of the
four roots to be $0$, $1$, and $\infty$. The GT-system compatibility
conditions then state that the fourth root $\lambda(u_1,u_2)$ is independent
of $u_2$. Using transformations of the form $u_1\to q(u_1)$, we obtain
either $\lambda=u_1$ or $\lambda=\mathrm{const}$. The GT-system
compatibility conditions then imply that in the first case, Eqs.~\eqref{93}
and~\eqref{94} coincide with the formulas in Example~\ref{ex2}, where
$P(x)=x(x-1)$, while the second case cannot be realized.

In the case of multiple roots, we can reduce the polynomial $R(x)$ to one of
the canonical forms $R=0$, $R=1$, $R=x$, $R=x^2$, or $R=x (x-1)$. In all
these cases, Eqs.~\eqref{93} and~\eqref{94} coincide with the corresponding
equations in Example~\ref{ex1}.

Below, we consider the general position case $R(x)=x(x-1)(x-u_1)$ and the
most degenerate case $R(x)=0$.

Because we use transformations~\eqref{95} when reducing $R$ to a canonical
form, the functions $g_k$ in system~\eqref{87} change their form from
polynomial to rational functions with the denominator $(p-\psi)^{k-1}$. To
construct them, we must describe all possible fractional-rational functions
$g_2$. Using a transformation of the form $\bar u_2=\sigma(u_1,u_2)$ in the
general position case, we can bring every such function $g_2$ to one of the
functions
\begin{align*}
&1.\quad g_2(p)=\frac{u_2(u_2-1)(p-u_1)}{u_1(u_1-1)(p-u_2)}\quad
\text{(the regular extension)},
\\[2mm]
&2.\quad g_2(p)=\frac1{p-u_1},
\\[2mm]
&3.\quad g_2(p)=\frac{u_1^{-\lambda}(u_1-1)^{\lambda-1}}{p-\lambda},\quad
\lambda=1,0,
\\[2mm]
&4.\quad g_2(p)=\frac{u_1-u_2}{u_1(u_1-1)}p+\frac{u_2-1}{u_1-1}.
\end{align*}
The regular extension, case~1, is invariant under the discrete
automorphisms described in Example~\ref{ex2}. The other three cases are
equivalent, and we can consider case~4 as an example.

We now consider the general position case~1. The next step in the
classification is finding fractional-rational GT families with the
coefficients depending on $u_1$ and $u_2$. Such a general form family is
determined by formulas~\eqref{50} and~\eqref{51}, where $n=2$ and $k=1$. But
we have the additional constraint that a zero of the denominator must
coincide with a zero of the numerator of $g_2$, i.e., it must be equal
$u_2$. It is easy to verify that this condition is equivalent to $s_2=0$ and
$h_{1,u_2}=h_{2,u_2}=0$ in system~\eqref{47},~\eqref{48}. The last condition
implies that $h_1(u_1)$ and $h_2(u_1)$ are linearly independent solutions of
the standard hypergeometric equation
\begin{equation}
u(u-1)h(u)''+[s_1+s_3-(s_3+s_4+2s_1)u]h(u)'+s_1(s_1+s_3+s_4+1)h(u)=0.
\label{96}
\end{equation}
Without restricting the generality, we can choose
$$
h_3(u_1,u_2)=\int_0^{u_2}(t-u_1)^{s_1}t^{s_3}(t-1)^{s_4}\,dt.
$$
As a result, we obtain
\begin{equation}
F(p,u_1,u_2)=\frac{f_1(u_1,u_2)p-f_2(u_1,u_2)}{p-u_2},
\label{97}
\end{equation}
where
\begin{align*}
&f_1=\frac{u_2(u_2-1)h_1h_{3,u_2}+u_1(u_1-1)(h_1h_{3,u_1}-h_3h_1')}
{u_1(u_1-1)(h_1h_2'-h_2h_1')},
\\[2mm]
&f_2=\frac{u_1u_2(u_2-1)h_1h_{3,u_2}+u_2u_1(u_1-1)(h_1h_{3,u_1}-h_3h_1')}
{u_1(u_1-1)(h_1h_2'-h_2h_1')}.
\end{align*}
We note that $h_1h_2'-h_2h_1'=\mathrm{const}\cdot(u_1-1)^{s_1+s_4}
u_1^{s_1+s_3}$.

For special values of the parameters, we can solve the hypergeometric
equation in elementary functions and find $F$ explicitly. For example, we
have
$$
F=\frac{(u_2-u_1)^{s_1+1}u_2^{s_3+1}(u_2-1)^{-1-s_1-s_3}}{p-u_2},
$$
for $s_4=-2-s_1-s_3$ and
$$
F=\frac{(p-1)(u_2-u_1)^{s_1+1}u_2^{s_3+1}(u_1-1)^{-1-s_1}}{p-u_2}
$$
for $s_4=0$.

Further, we must find the functions $g_3,g_4,\dots$ in system~\eqref{87}.
These functions are defined up to an arbitrary transformation~\eqref{84},
where $\alpha=3,4,\dots$. To fix them more or less rigidly, it is convenient
to assume that the functions $g_3,g_4,\dots$ are linear in $u_i$, $i>2$
(cf.~\eqref{90}). In particular, we can choose
\begin{align*}
&g_3(p)=-\frac{(u_1-u_2)(u_2-1)p}{u_1(u_1-1)(p-u_2)^2},
\\[2mm]
&g_i(p)=\frac{(i-3)(u_1-u_2)(u_2-1)pu_i}{u_1(u_1-1)(p-u_2)^2}-
\frac{(u_1-u_2)^{i-3}(u_2-1)^2p(p-u_1)(p-1)^{i-4}}
{u_1(u_1-1)^{i-2}(p-u_2)^{i-1}}-{}
\\[2mm]
&\phantom{g_i(p)={}}-\sum_{s=1}^{i-4}\frac{(i-s-2)(u_1-u_2)^s(u_2-1)^2
p(p-u_1)(p-1)^{s-1}u_{i-s}}{u_1(u_1-1)^{s+1}(p-u_2)^{s+2}},\quad i>3.
\end{align*}

The coefficients $\phi_{i,j}$ of the corresponding chain~\eqref{83} can be
determined from the relations
\begin{equation}
\begin{aligned}
&F(p)=\phi_{1,1}+\phi_{1,2}g_2,\qquad
F(p)g_2=\phi_{2,1}+\phi_{2,2}g_2+\phi_{2,3}g_3,
\\[2mm]
&F(p)g_3=\phi_{3,1}+\phi_{3,2}g_2+\phi_{3,3}g_3+\phi_{3,4}g_4,\qquad\dots,
\end{aligned}
\label{98}
\end{equation}
where $F$ is given by formula~\eqref{97}. These relations are equivalent to
an infinite triangular system of linear algebraic equations. Solving it, we
obtain
\begin{equation}
\begin{aligned}
&\phi_{1,1}=\frac{f_1u_1-f_2}{u_1-u_2},&\qquad
&\phi_{1,2}=-\frac{u_1(u_1-1)(f_1u_2-f_2)}{u_2(u_2-1)(u_1-u_2)},
\\[2mm]
&\phi_{2,1}=\frac{(u_2-1)(f_1u_2-f_2)}{(u_1-1)(u_1-u_2)},&\qquad
&\phi_{2,2}=\frac{f_2u_1-f_1u_2^2}{u_2(u_1-u_2)},
\\[2mm]
&\phi_{2,3}=f_1u_2-f_2,&\qquad &\ldots.
\end{aligned}
\label{99}
\end{equation}
All the functions $\phi_{i,j}$, $i>2$, are linear combinations of $f_1$ and
$f_2$ with rational coefficients with the same denominator
$(u_2-1)(u_1-u_2)$ and a numerator depending on $u_1,\dots,u_{i+1}$.

We next consider the most degenerate case $R(x)=0$. In this case, the GT
system is holomorphic on the diagonal. According to conjecture~\ref{gyp1}
(see Appendix~\ref{pr3}), precisely the models related to such GT systems
admit multidimensional generalizations.

We can easily verify that in the case $R(x)=0$, the triangular system is
equivalent to the system
$$
\ptl_ip_j=0,\qquad\ptl_i\ptl_ju_1=0,\qquad
\ptl_iu_k=p_i^{k-1}u_1,\quad k=2,3,\ldots.
$$
Automorphisms of this system are generated by the transformations
\begin{equation}
\begin{alignedat}2
&p_j\to p_j,&\qquad&u_i\to\nu u_i+\gamma_i,
\\[2mm]
&p_j\to ap_j+b,&\qquad
&u_i\to a^{i-1}u_i+(i-1)a^{i-2}bu_{i-2}+\dots+b^{i-1}u_1,
\end{alignedat}
\label{100}
\end{equation}
where $j=1,\dots,N$, $i=1,2,\dots$. The corresponding GT family is
$F(p)=A(u_1,u_2)p+B(u_1,u_2)$. The coefficients $A(x,y)$ and $B(x,y)$ are
found from the semi-Hamiltonian condition, which is equivalent to the system
of differential equations
\begin{equation}
\begin{alignedat}3
&AB_{yy}=A_yB_y,&\qquad&AB_{xy}=A_yB_x,&\qquad&AB_{xx}=A_xB_x,
\\[2mm]
&AA_{yy}=A_y^2,&\qquad&AA_{xy}=A_xA_y,&\qquad&AA_{xx}=A_x^2+A_xB_y-A_yB_x.
\end{alignedat}
\label{101}
\end{equation}
We can easily solve this system in elementary functions. For each solution
of it, formulas~\eqref{98} determine the corresponding integrable
chain~\eqref{83}.

Formulas~\eqref{101} imply that $F$ can depend on $u_2$ in two different
ways:
\begin{enumerate}
\item[1.]$F(p,u_1,u_2)=e^{\lambda u_2}(a(u_1)p+b(u_1))$ in the general
position case, or

\item[2.]$F(p,u_1,u_2)=a(u_1)p+\lambda u_2+b(u_1)$.
\end{enumerate}
In the first case, we in turn have two possibilities: $b'\ne0$ and $b'=0$.
We have
$$
a=\sigma',\qquad b=k_1\sigma,\qquad
\sigma(x)=c_1e^{\mu_1 x}+c_2e^{\mu_2 x},\quad
\text{where }c_1c_2(\lambda k_1-\mu_1\mu_2)=0,
$$
and
$$
b=c_1,\qquad a(x)=c_2e^{\mu x}+c_3,\quad
\text{where }c_2(c_1\lambda-c_3\mu)=0,
$$
for $b'=0$. The same variants in case~2 correspondingly give
$$
a=\sigma',\qquad b=k_1\sigma,\qquad\sigma(x)=c_1+c_2x+c_3e^{\mu x},
\quad\text{where }c_3(\lambda -c_2\mu)=0,
$$
and
$$
b=c_1,\qquad a(x)=c_2e^{\mu x}+c_3,\quad\text{where }c_2(\lambda-c_3\mu)=0.
$$
It is easy to see that in the general position case,
transformation~\eqref{100} reduces the function $F$ to the form
$$
F(p)=e^{u_2+u_1}(p-1)+e^{u_2-u_1}(p+1).
$$
The corresponding integrable chain is
\begin{equation}
u_{k,t}=(e^{u_2+u_1}+e^{u_2-u_1})u_{k+1,x}+(e^{u_2-u_1}-e^{u_2+u_1})u_{k,x},
\quad k=1,2,\ldots.
\label{102}
\end{equation}
This chain has an infinite hierarchy of commuting flows as usual. For
instance, the next flow is given by
\begin{align*}
u_{k,\tau}={}&(e^{u_2+u_1}+e^{u_2-u_1})u_{k+2,x}+
(u_3-u_1)(e^{u_2+u_1}+e^{u_2-u_1})u_{k+1,x}+{}
\\[2mm]
&{}+(e^{u_2+u_1}(u_1-u_3-1)+e^{u_2-u_1}(u_3-u_1-1))u_{k,x},\quad
k=1,2,\ldots.
\end{align*}
In case~2 with $c_3=\lambda=0$ and $k_1=1$, we obtain the chain
\begin{equation}
u_{k,t}=u_{k+1,x}+u_1u_{k,x},\quad k=1,2,\ldots,
\label{103}
\end{equation}
which is equivalent to universal hierarchy chain~\cite{48}.
Chain~\eqref{103} is a degeneration of the chain
\begin{equation}
u_{k,t}=u_{k+1,x}+u_2u_{k,x},\quad k=1,2,\ldots.
\label{104}
\end{equation}

Following~\cite{15},~\cite{41}, we can easily construct
$(2{+}1)$-dimensional generalizations of the $(1{+}1)$-dimensional chains
described above. Namely, some families of the obtained functions $F$ depend
linearly on two parameters. We let $\gamma_1$ and $\gamma_2$ denote these
parameters. The corresponding integrable chain
$$
u_{k,t}=\gamma_1(\phi_{k,1}u_{1,x}+\dots+\phi_{k,k+1}u_{k+1,x})+
\gamma_2(\psi_{k,1}u_{1,x}+\dots+\psi_{k,k+1}u_{k+1,x})
$$
is also linear in $\gamma_1$ and $\gamma_2$. We claim that the
$(2{+}1)$-dimensional chain
\begin{equation}
u_{k,t}=(\phi_{k,1}u_{1,x}+\dots+\phi_{k,k+1}u_{k+1,x})+
(\psi_{k,1}u_{1,y}+\dots+\psi_{k,k+1}u_{k+1,y})
\label{105}
\end{equation}
is integrable from the standpoint of the hydrodynamic reduction method. We
can easily describe these reductions in each actual case. For example, we
have
$$
F(p)=\gamma_1e^{u_2+u_1}(p-1)+\gamma_2e^{u_2-u_1}(p+1)
$$
in the general position case. The corresponding $(2{+}1)$-dimensional
chain~\eqref{105} is
$$
u_{k,t}=e^{u_2+u_1}(u_{k+1,x}-u_{k,x})+e^{u_2-u_1}(u_{k+1,y}+u_{k,y}),\quad
k=1,2,\ldots.
$$
Its hydrodynamic reductions are given by
$$
r^i_t=(e^{u_2+u_1}(p_i-1)+e^{u_2-u_1}q_i(p_i+1))r^i_x,\quad r^i_y=q_ir^i_x,
$$
where
\begin{equation}
\begin{aligned}
&\ptl_iu_k=p_i^{k-1}u_1,\quad k=1,2,\dots,\qquad
\ptl_i\,\ptl_ju_1=0,\qquad\ptl_ip_j=0,
\\[2mm]
&\ptl_iq_j=\frac{(q_i-q_j)[(p_i+1)(p_j-1)e^{u_1}+q_j(p_i-1)(p_j+1)e^{-u_1}]}
{(p_i-p_j)(e^{u_1}+q_i e^{-u_1})}\ptl u_1.
\end{aligned}
\label{106}
\end{equation}
Gibbons--Tsarev system~\eqref{106} with two-dimensional fibers (see
Appendix~\ref{pr3}) deserves a separate study.

We now consider one of the degenerate cases $F=\gamma_1e^{u_1}p+
\gamma_2(p+u_2)$. The corresponding $(2{+}1)$-dimensional integrable
generalization of chain~\eqref{104} is
$$
u_{k,t}=e^{u_1}u_{k+1,x}+u_{k+1,y}+u_2u_{k,y},\quad k=1,2,\ldots.
$$
The hydrodynamic reductions are described by
$$
r^i_t=[e^{u_1}p_i+q_i(p_i+u_2)]r^i_x,\qquad r^i_y=q_ir^i_x,
$$
where
\begin{align*}
&\ptl_iu_k=p_i^{k-1}u_1,\quad k=1,2,\dots,\qquad
\ptl_i\ptl_ju_1=0,\qquad\ptl_ip_j=0,
\\[2mm]
&\ptl_iq_j=\frac{(q_i-q_j)(p_iq_j+e^{u_1}p_j)}
{(p_i-p_j)(q_i+e^{u_1})}\ptl u_1.
\end{align*}
We note that the GT system for this $(2{+}1)$-dimensional chain is
equivalent to GT system~\eqref{106}.

\begin{gyph}
\label{gyp2}
Every integrable chain~\eqref{105} can be generated by a two-dimensional
linear family of solutions of system~\eqref{101} using the construction
described above.
\end{gyph}

\section{List of unsolved problems}
\label{pr5}

Here, we formulate several unsolved problems that from our standpoint seem
important for better understanding the nature of the GT system.

\begin{task}
\label{taskpr51}
To find all solutions of system~\eqref{32} of functional equations up to an
equivalence.
\end{task}

\begin{task}
\label{taskpr52}
To describe all two-field extensions of system~\eqref{33}.
\end{task}

\begin{task}
\label{taskpr53}
To describe all solutions of Eq.~\eqref{25} for GT
system~\eqref{45},~\eqref{46} (at least for small $n$).
\end{task}

\begin{task}
\label{taskpr54}
To construct solutions of Eq.~\eqref{25} and the pseudopotential in the case
of the regular GT-system extension in Example~\ref{ex4}.
\end{task}

\subsection*{Acknowledgments}
The authors are grateful to E.~V.~Ferapontov, who several years ago drew
their attention to interesting unsolved problems in the framework of the
hydrodynamic reduction method. The authors thank F.~F.~Voronov,
B.~A.~Dubrovin, I.~M.~Krichever, O.~I.~Mokhov, M.~V.~Pavlov,
V.~P.~Spiridonov, B.~L.~Feigin, E.~V.~Ferapontov, and A.~B.~Shabat for the
useful discussions. One of the authors (V.~V.~S.)~is grateful to Brock
University for the hospitality.

\smallskip

This work was supported in part by the Russian Foundation for Basic Research
(Grant Nos.~08-01-464?? and 09-01-92442) and by the Program for Supporting
Leading Scientific Schools (Grant No.~NSh-3472.2008.2).

\end{document}